\documentclass[a4paper,11pt]{article}
\pdfoutput=1 

\usepackage{jcappub} 

\usepackage[T1]{fontenc} 

\usepackage{amsmath}
\usepackage{graphicx,epsf,hyperref}

\usepackage{color}
\usepackage{xcolor}

\def\be{\begin{equation}}
\def\ee{\end{equation}}
\def\bea{\begin{eqnarray}}
\def\eea{\end{eqnarray}}
\def\lb{\left(}
\def\rb{\right)}
\def\lbs{\left[}
\def\rbs{\right]}
\def\lbc{\left\{}
\def\rbc{\right\}}
\def\p{\partial}
\def\n{\nabla}

\def\T{\mathcal{T}}
\def\Hh{\mathcal{H}}

\newcommand\itbf[1]{{\textit{\textbf{#1}}}}
\def\cs2{c_{\rm{s}}^2}
\def\U0{{\bar U_0}}
\def\12{\frac{1}{2}}

\def\U{{\Upsilon}}

\title{Magnetogenesis from isocurvature initial conditions}

 \author{Pedro Carrilho}
 \author{and Karim A. Malik}
 \affiliation{Astronomy Unit, School of Physics and Astronomy, Queen Mary University of London,\\
Mile End Road, London, E1 4NS, UK}

\emailAdd{p.gregoriocarrilho@qmul.ac.uk}
\emailAdd{k.malik@qmul.ac.uk}

\abstract{The generation of magnetic fields is a natural consequence of the existence of vortical currents in the pre-recombination era. This has been confirmed in detail for the case of adiabatic initial conditions, using second-order Boltzmann solvers, but has not been fully explored in the presence of isocurvatures. In this work, we use a modified version of the second-order Boltzmann code SONG to compute the magnetic field generated by vortical currents for general initial conditions. A mild enhancement of the generated magnetic field is found in the presence of general isocurvature modes, when compared to the adiabatic case. A particularly interesting case is that of the compensated isocurvature mode, for which the enhancement increases by several orders of magnitude due to the observationally allowed large amplitude of those modes. We show in this particular case how these compensated modes can influence observables at second order, such as the magnetic fields, and produce interesting effects which may be used to constrain these modes in the future.}

\begin{document}
\maketitle
\flushbottom

\section{Introduction}
\label{sec:intro}
Magnetic fields have been observed at all scales in the Universe. In galaxies, the field strengths range between a few to tens of $\mu$Gauss, correlated on kpc scales~\cite{Bernet:2008qp}. Magnetic fields of similar amplitude have also been measured in galaxy clusters, ordered on scales up to the Mpc~\cite{Feretti:2012vk}, and they have also been detected in superclusters~\cite{Xu:2005rb} and filaments~\cite{2010A&A...511L...5G} at smaller strengths. Even the voids of the large-scale structure are expected to host magnetic fields at the level of $10^{-15}$ G. These are estimated from observations of TeV blazars, whose emitted gamma-rays are expected to produce electron-positron cascades when interacting with background light. Upon inverse Compton scattering with the CMB, the charged particles are converted into GeV gamma-rays, which should reach observers on Earth, unless a magnetic field deflects the cascade away from the line-of-sight~\cite{Durrer:2013pga}. The lack of observation of these GeV halos has been used by the FERMI satellite to put \emph{lower} bounds on the void magnetic field strength of $10^{-13}$ --- $10^{-15}$ G, depending on assumptions about the gamma-ray jet life-time~\cite{Biteau:2018tmv}.\footnote{Note, however, that these results have been put into question by Broderick et al. in Ref.~\cite{Broderick:2018nqf}. In this work, TeV sources are observed off-axis, for which the authors argue that GeV gamma-rays should be detected. Their non-detection instead implies an \emph{upper} bound on the strength of inter-galactic magnetic fields of $10^{-15}$~G.}

Galactic magnetic fields are explained by current theories via a dynamo mechanism, activated in the final stages of gravitational collapse. This mechanism amplifies a pre-existing magnetic field until it saturates at the equipartition level of $\mu$G, at which we observe it today. The origin of the magnetic seed required to activate this mechanism is so far unknown and its size is difficult to estimate from these measurements, given that our observations only probe the saturated magnetic field and modelling of the dynamo is non-trivial~\cite{Brandenburg:2004jv}. However, using simplified arguments and assuming the efficiency of the dynamo is well understood, lower bounds can be found, ranging from the optimistic $10^{-30}$ G, assuming equipartition was only reached today~\cite{Davis:1999bt}, to $10^{-15}$ G when this saturation occurs already at redshift $z\sim2$~\cite{Durrer:2013pga}. 

Both magnetic fields in voids and those in galaxies and other structures could have a common origin. Astrophysical processes could explain these fields, in particular if generated during the complex stages of structure collapse, in which non-linear dynamics plays a role. Another alternative is that these seed magnetic fields could be generated primordially, before structure formation takes place. Should that be the case, this primordial magnetic field would affect the Cosmic Microwave Background (CMB) through many different effects. Since these effects have remained undetected, Planck~\cite{Ade:2015cva} has placed upper bounds of approximately $10^{-9}$ G on the amplitude of magnetic fields at Mpc scales. Upper limits have also been placed on the total, integrated, magnetic field, at the level of $10^{-12}$ G~\cite{Jedamzik:2018itu}. Faraday rotation measures have also been used to place constraints on extra-galactic magnetic fields at the $n$G level, which are independent on the origin and generation mechanism of magnetic fields~\cite{Blasi:1999hu,Pshirkov:2015tua}.

Many models for the generation of primordial magnetic fields exist~\cite{Grasso:2000wj, Widrow:2002ud, Durrer:2013pga}, which differ greatly in terms of the epoch of magnetogenesis as well as in the amount of exotic physics necessary. Inflationary models can generate appreciable magnetic fields on very large scales, but they require the introduction of new physics to break the conformal invariance of electromagnetism~\cite{Turner:1987bw,Dimopoulos:2001wx,Ashoorioon:2004rs,Ferreira:2013sqa,Caprini:2017vnn}. Phase transitions, such as the GUT, electroweak or QCD transitions can generate considerable amplitudes for the magnetic fields~\cite{Vachaspati:1991nm,Kamada:2018tcs}, but their correlation length is always very small, unless their magnetic helicity is substantial~\cite{Caprini:2009pr}. 

A more conservative alternative is the generation of vortical currents in the early Universe, after electron-position annihilation~\cite{1970MNRAS.147..279H,Matarrese:2004kq, Gopal:2004ut,Takahashi:2005nd,Ichiki:2006cd,Siegel:2006px, Ichiki:2007hu,Kobayashi:2007wd, Maeda:2008dv, Fenu:2010kh,Maeda:2011uq, Nalson:2013jya, Saga:2015bna, Fidler:2015kkt}. This mechanism is always present, as it only requires standard electromagnetism and the dynamics of fluctuations in the baryon-photon plasma. However, magnetic fields generated through this mechanism have a rather small amplitude, since the effect appears only at second order in cosmological perturbations and is suppressed by the tight coupling of baryons and photons. The most recent calculation of the spectrum of magnetic fields generated through this mechanism are detailed in Ref.~\cite{Fidler:2015kkt} and found an amplitude of order $10^{-29}$ G on Mpc scales at $z=0$. 

Most previous works have assumed that the initial conditions for cosmological fluctuations are adiabatic. However, it has been suggested by Maeda et al. in Ref.~\cite{Maeda:2011uq} and Nalson et al. in Ref.~\cite{Nalson:2013jya}, that non-adiabatic fluctuations could enhance the magnetic field produced via this mechanism. Those conclusions are based on analytical calculations, which do not include all effects due to recombination and therefore require further confirmation. In this paper, we use the Boltzmann solver SONG\footnote{https://github.com/coccoinomane/song} \cite{pettinari:2013a,pettinari:2015a} with isocurvature initial conditions to compute the enhancement of the spectrum of magnetic fields created by these non-adiabatic modes. We are informed by the most recent results from Planck regarding the possible amplitude of the spectrum of primordial isocurvature modes and their spectral index~\cite{Akrami:2018odb}, but also expand on those possibilities, to better understand the effect of generic isocurvatures. We are also particularly interested in exploring the existence of a compensated isocurvature mode~\cite{2010ApJ...716..907H,Grin:2011tf,Grin:2013uya,He:2015msa}. This mode is an anti-correlated mixture of the baryon and dark matter (DM) density isocurvatures, in which those modes compensate each other to avoid the existence of a matter isocurvature mode. This fact implies that its primordial amplitude is very difficult to constrain. However, this mode does generate an observable effect in the CMB by modulating the background baryon-to-DM ratio, which has allowed Planck to estimate its amplitude to be six orders of magnitude larger than that of the adiabatic mode~\cite{Akrami:2018odb}. In this work, we also investigate the effects of this mode on magnetic field generation, since its large amplitude should allow for a substantial enhancement of the magnetic power spectrum.

This paper is organized in the following way: in Section~\ref{sec:mag}, we review the mechanism responsible for generating magnetic fields and show its evolution equations; in Section~\ref{sec:res} we show our numerical results for the spectrum of magnetic fields generated with different types of isocurvature initial conditions, including the compensated isocurvature mode. Finally, in Section~\ref{sec:conc}, we discuss our findings and explain the relevance of our results for the study of the cosmological generation of magnetic fields, as well as for the study of the early Universe in general. In Appendix \ref{sec:ini}, we detail the initial conditions for the vector degrees of freedom required to initialize the numerical evolution, following the techniques laid out in Ref.~\cite{Carrilho:2018mqy}.

\section{Magnetogenesis from vortical currents}
\label{sec:mag}

The mechanism for magnetogenesis studied here acts in the early Universe, starting from around the end of electron-positron annihilation and being active almost until today. During this entire stage, the species of interest are photons ($\gamma$), electrons ($e$), and protons ($p$), which comprise the tightly coupled baryon-photon plasma, prior to recombination. Additionally, cold dark matter (c) and neutrinos ($\nu$) are also present, but do not affect the generation of magnetic fields directly. We assume here Einstein's general relativity to describe the geometry in which these species live. The metric tensor, $g_{\alpha\beta}$, evolves according to the Einstein field equations
\be 
R^{\alpha\beta}-\frac12 g^{\alpha\beta}R=8\pi G T^{\alpha\beta}\,,
\ee
in which $R^{\alpha\beta}$ is the Ricci tensor, $R$ is the Ricci scalar, $G$ is Newton's constant and $T^{\alpha\beta}$ is the stress-energy tensor, which is given by the sum of the stress-energy tensors of all species (s),
\be
T^{\alpha\beta}=\sum_s T_s^{\alpha\beta}\,,
\ee
which are given by
\begin{align}
&T_{\text{c}}^{\alpha\beta}=\rho_{\text{c}} u_{\text{c}}^\alpha u_{\text{c}}^\beta\,,\\
&T_e^{\alpha\beta}=\rho_e u_e^\alpha u_e^\beta\,,\\
&T_p^{\alpha\beta}=\rho_p u_p^\alpha u_p^\beta\,,\\
&T_\gamma^{\alpha\beta}=\frac43\rho_\gamma u_\gamma^\alpha u_\gamma^\beta+\frac13 \rho_\gamma g^{\alpha\beta}+\pi_\gamma^{\alpha\beta}\,,\\
&T_\nu^{\alpha\beta}=\frac43\rho_\nu u_\nu^\alpha u_\nu^\beta+\frac13 \rho_\nu g^{\alpha\beta}+\pi_\nu^{\alpha\beta}\,.
\end{align}
We have here defined the energy densities of each species as $\rho_s$, the 4-velocity vectors as $u_s^\alpha$, and the anisotropic stress tensors for photons and neutrinos as $\pi_\gamma^{\alpha\beta}$ and $\pi_\nu^{\alpha\beta}$, respectively. We are treating electrons, protons and dark matter as pressureless perfect fluids, as is clear by the absence of pressure and anisotropic stress in the their stress-energy tensors. Photons and neutrinos are assumed to be relativistic species with $P_s=\rho_s/3$. We have written all species in their respective energy frames, as shown by the lack of an energy flux term, $q^\mu$ in the expressions above.

In addition to these species, an electromagnetic field is present, which is described by the Faraday tensor, $F_{\mu\lambda}$. Using a normalised time-like 4-vector field, $u^\alpha$, to represent a set of observers, one may then define an electric field, $E^\mu$, and a magnetic field, $B^\mu$, via
\be
E^\mu=F^{\mu\lambda}u_\lambda\,, \ B^\mu=u_\alpha \epsilon^{\alpha\mu\lambda\beta}F_{\lambda\beta}\,,
\ee
in which $\epsilon^{\alpha\mu\lambda\beta}$ is the totally antisymmetric tensor, with $\epsilon_{0123}=\sqrt{-g}$, with $g$ the determinant of the metric.

The particle species evolve according to their Boltzmann equations coupled to the Einstein and Maxwell equations. For the purposes of this paper, it suffices to display here only the equations for the first two multipoles of the distribution functions, which can be written in terms of the divergence of the stress-energy tensors of the different species:
\begin{align}
&\n_\alpha T_c^{\alpha\mu}=0\,,\\
&\n_\alpha T_e^{\alpha\mu}=F^\mu_{\ \lambda} j^\lambda_e+C^\mu_{e\gamma}+C^\mu_{ep}\\
&\n_\alpha T_p^{\alpha\mu}=F^\mu_{\ \lambda} j^\lambda_p+C^\mu_{p\gamma}-C^\mu_{ep}\\
&\n_\alpha T_\gamma^{\alpha\mu}=-C^\mu_{p\gamma}-C^\mu_{e\gamma}\,,\\
&\n_\alpha T_\nu^{\alpha\mu}=0\,.
\end{align}
The electric currents $j_s^\mu$ are given by
\be
j_s^\mu=q_s n_s u_s^\mu\,,
\ee
with $q_s$ the charge of the particles in question, being equal to the fundamental electronic charge, $e$, for protons and $-e$ for electrons. The symbol $n_s$ denotes the number density of the species in question as seen by an observer in the $u_s$ frame. $C_{sr}^\mu$ are the collision terms for the interactions between the species $s$ and $r$, which will be detailed below.

The electromagnetic field obeys Maxwell's equations
\be
\n_\lambda F^{\mu\lambda}=j_e^\mu+j_p^\mu\,,\ \ \n_{[\alpha}F_{\mu\lambda]}=0\,,
\ee
which can be written in terms of the electric and magnetic fields as
\begin{align}
\n_\alpha B^\alpha+u^\alpha u^\beta\n_\beta B_\alpha &=-\epsilon_{\alpha\beta\mu\lambda}E^\alpha u^\beta\n^\lambda u^\mu\,,\\
\n_\alpha E^\alpha+u^\alpha u^\beta\n_\beta E_\alpha &=\epsilon_{\alpha\beta\mu\lambda}B^\alpha u^\beta\n^\lambda u^\mu+\varrho\,,\\
u^\alpha\n_\alpha B^\mu+B^\mu\n_\alpha u^\alpha-B^\alpha\n_\alpha u^\mu+& u^\alpha u^\beta u^\mu\n_\beta B_\alpha \nonumber\\ 
&=-\epsilon^{\mu\alpha\beta\lambda} u_\alpha\n_\lambda E_\beta-h^{\lambda\mu}\epsilon_{\lambda\sigma\alpha\beta}\n^\sigma u^\alpha E^\beta\,,\label{FaradayFull}\\
u^\alpha\n_\alpha E^\mu+E^\mu\n_\alpha u^\alpha-E^\alpha\n_\alpha u^\mu+& u^\alpha u^\beta u^\mu\n_\beta E_\alpha \nonumber\\
&=\epsilon^{\mu\alpha\beta\lambda} u_\alpha\n_\lambda B_\beta+h^{\lambda\mu}\epsilon_{\lambda\sigma\alpha\beta}\n^\sigma u^\alpha B^\beta-J^\mu\,,
\end{align}
in which $h^{\mu\lambda}=g^{\mu\lambda}+u^\mu u^\lambda$ is the projection tensor to the space perpendicular to the frame $u^\mu$ and $\varrho$ and $J^\mu$ are projections of $j^\mu=j^\mu_e+j^\mu_p$ given by
\be
\varrho=-u_\mu j^\mu\,,\ \ J^\mu=h^\mu_{\lambda}j^\lambda\,.
\ee

We now expand these equations around the flat Friedmann--Lema\^{i}tre--Robertson--Walker (FLRW) spacetime, up to second order in cosmological fluctuations and we show them in Poisson gauge, for which the line element simplifies to
\be
\text{d}s^2=a(\eta)^2\lbs-(1+2\phi)\text{d}\eta^2-2S_i\text{d}x^i\text{d}\eta+(1-2\psi)\delta_{ij}\text{d}x^i\text{d}x^j\rbs\,,
\ee
which we have written in terms of conformal time and have performed a scalar-vector-tensor decomposition. We denote by $a(\eta)$ the scale factor, by $S_i$ the vector part of the shift --- the only non-zero vector potential present in the metric in this gauge --- by $\phi$ the perturbation to the lapse and by $\psi$ the curvature fluctuation in this gauge. These scalar potentials as defined in this gauge are equal to the two gauge invariant Bardeen potentials \cite{Bardeen:1980kt}. For simplicity, we have neglected the tensor mode as it will not be important for the magnetic field calculation and we will assume that the vector mode, $S_i$, is only non-zero at second order, as we are not considering primordial vector modes. Our normalization for second-order quantities includes a factor of $1/2$, so that the vector mode, being purely second order, is given by $S_i=\frac12S_i^{(2)}$.

Other variables are expanded and decomposed into scalars, vectors and tensors in the standard way, as given in Refs.~\cite{Malik:2008im,Carrilho:2015cma}. In particular, in the Poisson gauge, the 4-velocity is expanded as
\begin{align}
&u^{0}=a^{-1}\left(1-\phi+\frac32\phi^2+\frac12v_i v^i\right)\,,\\
&u^{i}=a^{-1}v^i=a^{-1}\lb v^{,i}+v^i_\text{v}\rb\,,
\end{align}
while the anisotropic stress is given by
\begin{align}
\pi_{00}=0,\ \ \ \pi_{i0}=-2\pi_{ij} v^j\,,\nonumber\\
\pi_{ij}=a^2\lbs\Pi_{ij}+\Pi_{(i,j)}+\Pi_{,ij}-\frac{1}{3}\delta_{ij}\n^2\Pi\rbs\,.
\end{align}
The electric and magnetic fields are also decomposed and expanded up to second order in perturbations. This results in
\begin{align}
E_0=- E_i v^i\,,\ E_i=a (E_{\text{v}\,i}+E_{,i})\,,\\
B_0=- B_i v^i\,,\ B_i=a (B_{\text{v}\,i}+B_{,i})\,.
\end{align}

The collision terms are given in Refs.~\cite{Fenu:2010kh,Maeda:2011uq,Saga:2015bna}. The interaction term for Coulomb scattering is proportional to the velocity difference between electrons and protons. This interaction between the charged species is very strong and completely dominates the dynamics at early times, giving rise to a tightly coupled fluid, in which the velocity fields of its constituents nearly match. This implies electrons and protons can be considered a single fluid of baryons with velocity, $v_{\text{b}}$, given by
\be
v_{\text{b}}^i=\frac{m_p v_p^i+m_e v_e^i}{m_p+m_e}\,.
\ee
As shown in Ref.~\cite{Fenu:2010kh}, for the reason above, the collision term between electrons and protons does not enter the calculation of the magnetic field and, for brevity, we do not show it here. 

The momentum transfer rates for the Compton/Thomson interactions are obtained by projecting the collision terms for that interaction with $h_{\mu\lambda}$. For the charged species $r$, the momentum transfer rate is given by
\be
C^{r\gamma}_{\, i}=\frac43 \rho_\gamma \lb\frac{m_e}{m_r}\rb^2\kappa' (1+\delta_r+\delta_\gamma-2\psi) (v_{r\,i}-v_{\gamma\,i})+ \lb\frac{m_e}{m_r}\rb^2\kappa' \frac{1}{a^2}v_r^j\pi_{\gamma\, ij}\,,\label{ThomsonC}
\ee
where we have defined the interaction rate $\kappa'=-a n_e \sigma_T$, with $\sigma_T$ the Thomson cross section and $n_e$ the background number density of free electrons, which is equal to that of free protons and will often be denoted simply as $n$. We have also defined the density contrasts $\delta_s=\delta\rho_s/\rho_s$, in which $\rho_s$ is the background density of species $s$ and $\delta\rho_s$ is its density perturbation. The density contrast $\delta_r$ represents here only the free charged particles.

It is clear from Eq.~\eqref{ThomsonC} that the Compton interaction is far more effective for electrons than it is for protons, given their substantial mass difference. This gives rise to charge separation due to this imbalance. An electric field is thus generated, which is given by\footnote{This equation is derived in detail in Ref.~\cite{Fenu:2010kh}, in which an analysis of the different time-scales of the problem is included and the term shown here is concluded to be the dominant one.}
\be
\label{Elefield}
E_i=-\frac{1-\beta^3}{1+\beta}\frac{a\sigma_T}{e}\lb\frac{4}{3}\rho_\gamma(1+\delta_\gamma-2\psi)\Delta v_{\text{b}\gamma\, i}+\frac{1}{a^2}v_{\text{b}}^j\pi_{\gamma\, ij}\rb\,,
\ee
where we have introduced new notation for the mass ratio $\beta=m_e/m_p$ and the velocity difference $\Delta v_{\text{b}\gamma\, i}=v_{\text{b}\,i}-v_{\gamma\,i}$. As stated above, it is the mass difference between electrons and protons which gives rise to this electric field, as this would not be possible with $\beta=1$. It is for a similar reason that this electric field can only be generated after electron-positron annihilation, as, before that, the mass ratio of relevance is that of positrons and electrons, which is unity. 

A magnetic field can thus be generated via Faraday's law, Eq.~\eqref{FaradayFull}. Expanding that equation up to second order in fluctuations, one finds
\be
\label{Faradcoord}
\lb a B_i\rb'=-a\epsilon_{ij}^{\ \ k}\partial^j\lb (1+\phi+v')E_k\rb\,.
\ee
We see the frame-dependence of this field very clearly in the term with $v'$. This frame is often chosen to be a local inertial frame with the observer's 4-velocity, $u^\mu$, being aligned with one of the axis of a tetrad basis, $e_{\underline{a}}^\mu$. Should this alignment be such that $e_{\underline{0}}^\mu=u^\mu$, and in the Poisson gauge, we have $v=0$, $E_k=a (1-\psi) E_{\underline{k}}$ and $B_i=a B_{\underline{i}}$, where the underlined indices label the components of tensors in the tetrad basis.\footnote{This is the tetrad used in Ref.~\cite{Fenu:2010kh}, and the components of the basis vectors are presented in its Appendix B. Other choices may be made, such as $e^{\underline{0}}_\mu=u_\mu$, as is done in Refs.~\cite{pettinari:2015a, Carrilho:2018mqy} and in the Boltzmann solver SONG, which we use here. It can be shown that both choices give the same evolution equation for the magnetic field.} In that case, Eq.~\eqref{Faradcoord} becomes
\be
\label{Faradtetrad}
\lb a^2 B_{\underline{i}}\rb'=-a^2\epsilon_{ij}^{\ \ k}\partial^j\lb (1+\phi-\psi)E_{\underline{k}}\rb\,,
\ee
which is the version used in Refs.~\cite{Fenu:2010kh,Fidler:2015kkt} and which we also use in our numerical studies below.

Different frame choices have been studied in Ref.~\cite{Fenu:2010kh} and the results show differences between the baryon frame and the fundamental frame, at early times and large scales, but no effect at $z=0$. One could also use an alternative frame, such as the energy frame, but the results are not expected to vary significantly, unless a frame is chosen with a very high velocity with respect to the frame comoving with the expansion. 

Another well-known aspect of this mechanism is that it does not occur at first order in fluctuations, in the absence of primordial vector modes. This is transparent in Eqs.~\eqref{Faradcoord} and~\eqref{Faradtetrad}, since the magnetic field is sourced by the curl of the electric field, which would only be non-zero at linear level in the presence of linear vectors.

Finally, we are able to write the evolution equation for the magnetic field in terms of variables commonly computed in a Boltzmann solver:
\be
\label{Faraddv}
\lb a^2 B_{\underline{i}}\rb'=a^2\frac{1-\beta^3}{1+\beta}\frac{\sigma_T}{e}\epsilon_{ijk}\partial^j\lb \frac43\rho_\gamma\lb\Delta v_{\text{b}\gamma\, \text{v}}^{\ \ k}+(\delta_\gamma+\phi-2\psi)\Delta v_{\text{b}\gamma}^{\ \ ,k}\rb+v_{\text{b}\,,l}D^{lk}\Pi_\gamma\rb\,,
\ee
with $D^{kl}=\p^k\p^l-\frac13\delta^{kl}\n^2$ and where we have written the right-hand-side using the scalar-vector-tensor decomposition. It is clear from Eqs.\eqref{Elefield} and \eqref{Faraddv} that the electromagnetic field depends on $\Delta v_{\text{b}\gamma}$ and $\Pi_\gamma$, both of which are strongly suppressed by the strong Thomson interactions in the early Universe. However, this tight-coupling becomes less effective around the time of recombination and an appreciable magnetic field is generated at that time.

To solve Eq.~\eqref{Faraddv}, one needs to solve the full system of linear scalar fluctuations, as well as the system of second-order vector perturbations. As mentioned above, we will use the Boltzmann solver SONG~\cite{pettinari:2013a,pettinari:2015a} to solve the vector equations numerically, which makes use of the linear solver CLASS~\cite{Lesgourgues:2011re,Blas:2011rf} to solve for the linear evolution.

\subsection{Isocurvature modes}

The solution of the Einstein-Boltzmann system requires the specification of initial conditions. Different choices of initial conditions would result in distinct solutions to the evolution equations, which can be probed with experimental data. An analysis of the linear Einstein-Boltzmann system~\cite{Bucher:1999re} has led to the classification of its solution space into regular (growing) and singular (decaying) modes. While decaying modes have also been studied~\cite{Amendola:2004rt}, growing modes are certain to exist and we focus only on those. These growing modes can further be decomposed into an adiabatic and four isocurvature modes, at first order~\cite{Bucher:1999re}, but only three isocurvature modes source growing solutions at second order~\cite{Carrilho:2018mqy}. We define entropy fluctuations of species $r,s$ as
\be
S_{rs}=\frac{\delta_r}{1+w_r}-\frac{\delta_s}{1+w_s}\,,
\ee
with $w_s=P_s/\rho_s$ the equation of state parameter for species $s$. The adiabatic mode is that for which all entropy fluctuations vanish initially.

The three isocurvature modes are defined instead by the initial vanishing of the curvature fluctuation $\zeta$, defined as 
\be
\zeta=-\psi-\frac{\delta}{3(1+w)}\,,
\ee
at first order. Each isocurvature mode is defined by having one of $S_{s\gamma}$ non-zero and are named the baryon, cold dark matter and neutrino density isocurvature modes, depending on which entropy fluctuation is non-zero. A more detailed definition of isocurvature modes is included in Ref.~\cite{Carrilho:2018mqy}, which is also consistently extended to second order.

Isocurvature modes can have an influence on the generation of magnetic fields. Maeda et al.~\cite{Maeda:2011uq} have shown that, in the presence of a baryon isocurvature mode ($S_{\text{b}\gamma}\neq0$), the evolution equation for the magnetic field, Eq.~\eqref{Faraddv}, can be written as 
\be
\label{FaraddvMaeda}
\lb a^2 B_{\underline{i}}\rb'=\frac{1-\beta^3}{1+\beta}\frac{4\sigma_T}{3e}a^2\rho_\gamma\lb C_\omega(\eta)\,\omega_i+C_S(\eta) \epsilon_{ijk}S_{\text{b}\gamma}^{\ \ ,j}\Delta v_{\text{b}\gamma}^{\ \ ,k}\rb\,,
\ee
where the tight coupling expansion~\cite{Pitrou:2010ai} was used up to first order in $\Hh/\kappa'$ to approximate the right-hand-side analytically. For that reason, this expression is only valid at early times, long before recombination. The symbols $C_\omega$ and $C_S$ represent time-dependent functions, which we do not specify, and $\omega_i$ is the vorticity of the total fluid. It is clear that without the baryon isocurvature mode, the second term in Eq.~\eqref{FaraddvMaeda} would vanish. The same is also true for the first term, since vorticity can only be generated in the presence of non-adiabatic pressure, as is well known~\cite{Christopherson:2009bt,Christopherson:2010ek,Christopherson:2010dw}. Therefore, a magnetic field cannot be generated at first order in tight coupling in the absence of this isocurvature mode, as argued by Maeda et al.~\cite{Maeda:2011uq}. A similar conclusion appears in Appendix D of Ref.~\cite{Fenu:2010kh} by Fenu et al., in which it is noted that a large suppression exists in the tight coupling approximation when only the adiabatic mode in considered. Similar arguments are also made in Ref.~\cite{Nalson:2013jya}. These conclusions point to the fact that a non-adiabatic mode can provide a large contribution to the source of the magnetic field, at least at sufficiently early times, when the tight coupling expansion is valid. The magnetic field spectrum has been computed analytically in this approximation, up to matter-radiation equality, by Maeda et al. and found to be larger than in the adiabatic case. However, this calculation ignores all effects occurring at a higher order in tight coupling and, especially, it does not take recombination into account, which is the moment in which the sources of the magnetic field are more important.

Another relevant aspect regarding isocurvatures is mode mixing. Given that the adiabatic mode is certain to exist, the presence of an isocurvature mode implies that the two modes will mix due to the non-linear nature of the evolution equations at second order. Beyond generating source terms which do not exist when each of the modes is considered individually, this coupling between modes is particularly important for the sources of the magnetic field, as they include cross products between gradients of scalar variables, i.e. $\epsilon_{ijk}\p^j A\, \p^k B$. At sufficiently early times, scalar variables are proportional to the initial value of $\zeta$ or $S_{r\gamma}$, depending on the modes being considered. If a single mode is present, the cross products vanish, as $\epsilon_{ijk}\p^j A \p^k B\propto\epsilon_{ijk}\p^j I \p^k I=0$. Evidently, the mixing of multiple modes avoids this issue, as one would have, for example, $\epsilon_{ijk}\p^j A \p^k B\propto\epsilon_{ijk}\p^j \zeta \p^k S_{\text{b}\gamma}\neq0$, unless the two modes are fully correlated or anti-correlated i.e. $S_{\text{b}\gamma}\propto\zeta$.

The arguments made here are valid at early times and large scales, but require more general investigation. This is what motivates us to explore the evolution of magnetic fields in the presence of isocurvature modes numerically using the Boltzmann solver SONG. This has required a modification of the publicly available version of that code to include isocurvature initial conditions. This was studied in Ref.~\cite{Carrilho:2018mqy} for scalar modes and we use the same techniques here to compute approximate initial solutions for vector modes for all isocurvature modes under consideration. The results for the initial conditions are shown in Appendix~\ref{sec:ini}. 

\section{Numerical results}
\label{sec:res}

\subsection{Set-up}

We are interested in computing the spectrum of the magnetic field defined as
\be
\left\langle B_{\underline{i}}\lb\textit{\textbf{k}},\eta\rb B^{\underline{i}}\lb\textit{\textbf{k}'},\eta\rb\right\rangle=(2\pi)^3 \delta^{(3)}\lb\textit{\textbf{k}}+\textit{\textbf{k}'}\rb P_B(k,\eta)\,.
\ee
It is convenient, for numerical reasons, to decompose the magnetic field with polarization vectors $e_{\pm}^{\underline{i}}$, perpendicular to the direction $\textit{\textbf{k}}$, such that $B^{\underline{i}}=B_+ e_{+}^{\underline{i}}+ B_- e_{-}^{\underline{i}}$. We write these magnetic field components in terms of transfer functions as
\be
B_\pm(\textit{\textbf{k}},\eta)=\int \frac{\text{d}^3k_1\text{d}^3k_2}{(2\pi)^3}\delta^{(3)}(\textit{\textbf{k}}-\textit{\textbf{k}}_1-\textit{\textbf{k}}_2)\mathcal{T}_{B_\pm}^{ab}(\textit{\textbf{k}},\textit{\textbf{k}}_1,\textit{\textbf{k}}_2,\eta)I_a(\textit{\textbf{k}}_1) I_b(\textit{\textbf{k}}_2)\,,\label{Btransfer}
\ee
in which $I_a$ denote the variables defining the mode under consideration, taking values in the set $\lbc\zeta,S_{\text{c}\gamma},S_{\text{b}\gamma},S_{\nu\gamma}\rbc$ for the adiabatic mode, cold dark matter, baryon and neutrino isocurvature modes, respectively. The indices $a,b,c,d$ represent those modes, taking four different values: $\zeta,\text{c},\text{b},\nu$ for their respective modes. The Einstein summation convention was employed for those indices. The spectra of the defining variables $I_a$ are defined as
\be
\left\langle I_{a}\lb\textit{\textbf{k}},\eta\rb I_b\lb\textit{\textbf{k}'},\eta\rb\right\rangle=(2\pi)^3 \delta^{(3)}\lb\textit{\textbf{k}}+\textit{\textbf{k}'}\rb P_{ab}(k)\,.
\ee
These spectra are parameterized by an amplitude $A_{ab}$ and a spectral index $n_{ab}$, so that they are given by
\be
P_{ab}(k)=A_{ab}\frac{2\pi^2}{k^3}\lb\frac{k}{k_*}\rb^{n_{ab}}\,,
\ee
in which we have denoted the pivot scale by $k_*$ and will choose it to be $k_*=0.05\ \text{Mpc}^{-1}$. For $a=b$, one has the standard power spectra, such as the power spectrum of the curvature perturbation $P_{\zeta\zeta}$ for the adiabatic mode, for which the amplitude is more commonly denoted $A_s$ and the spectral index is given by $n_{\zeta\zeta}=n_s-1$. The cases with $a\neq b$ represent the correlations between different modes, which is often parameterized by the correlation fraction
\be
\cos\Delta_{ab}=\frac{P_{ab}}{\sqrt{P_{aa} P_{bb}}}\,,
\ee
which, as the notation indicates, must obey $-1<\cos\Delta<1$. Given our power law parameterisation, this can be shown to imply that $n_{ab}=(n_{aa}+n_{bb})/2$.

The transfer functions obey the identities given by
\begin{align}
\T^{ab}(\textit{\textbf{k}},\textit{\textbf{k}}_1,\textit{\textbf{k}}_2,\eta)=\T^{ba}(\textit{\textbf{k}},\textit{\textbf{k}}_2,\textit{\textbf{k}}_1,\eta)\,,\\
\T^{ab}(-\textit{\textbf{k}},-\textit{\textbf{k}}_1,-\textit{\textbf{k}}_2,\eta)=\T^{ab}(\textit{\textbf{k}},\textit{\textbf{k}}_1,\textit{\textbf{k}}_2,\eta)\,.
\end{align}
Thus, one can show that, in this notation, the magnetic field spectrum is given by
\be
\label{SpcTra}
P_B(k,\eta)=4\int\frac{\text{d}^3q}{(2\pi)^3}\mathcal{T}^{ab}_{B_+}(\textit{\textbf{k}},\textit{\textbf{q}},\textit{\textbf{k}}-\textit{\textbf{q}},\eta)\mathcal{T}^{cd}_{B_+}(\textit{\textbf{k}},\textit{\textbf{q}},\textit{\textbf{k}}-\textit{\textbf{q}},\eta)P_{ac}(\textit{\textbf{q}})P_{bd}(\textit{\textbf{k}}-\textit{\textbf{q}})\,,
\ee
assuming Gaussian initial conditions and that helical magnetic fields are not generated~\cite{Saga:2015bna}.

We now present our results for the magnetic field generated via this mechanism in the presence of isocurvature modes. We begin by showing the results for the addition of single isocurvature modes and study how the amplitudes and spectral tilts of the different modes affect the magnetic field production via this mechanism. We then show the results obtained for a mixture of isocurvature modes which is particularly interesting --- the so-called compensated isocurvature mode. All cosmological parameters used that are not related to isocurvatures, are taken to be the best-fit values from Planck, as given in Ref.~\cite{Akrami:2018odb}.

\subsection{Magnetogenesis with single isocurvature mode}

We begin by showing results for the magnetic field power spectrum, $P_B$, generated when a single isocurvature mode is present in addition to the adiabatic mode. We assume, initially, that the isocurvature mode has a scale-invariant spectrum ($n_{aa}=0$) with the same amplitude as the adiabatic mode ($A_{aa}=A_s$), but uncorrelated with it ($\cos\Delta_{a\zeta}=0$). We will later change these assumptions to measure the importance of those parameters. 

\begin{figure}
    \centering
    \includegraphics[width=0.49\textwidth]{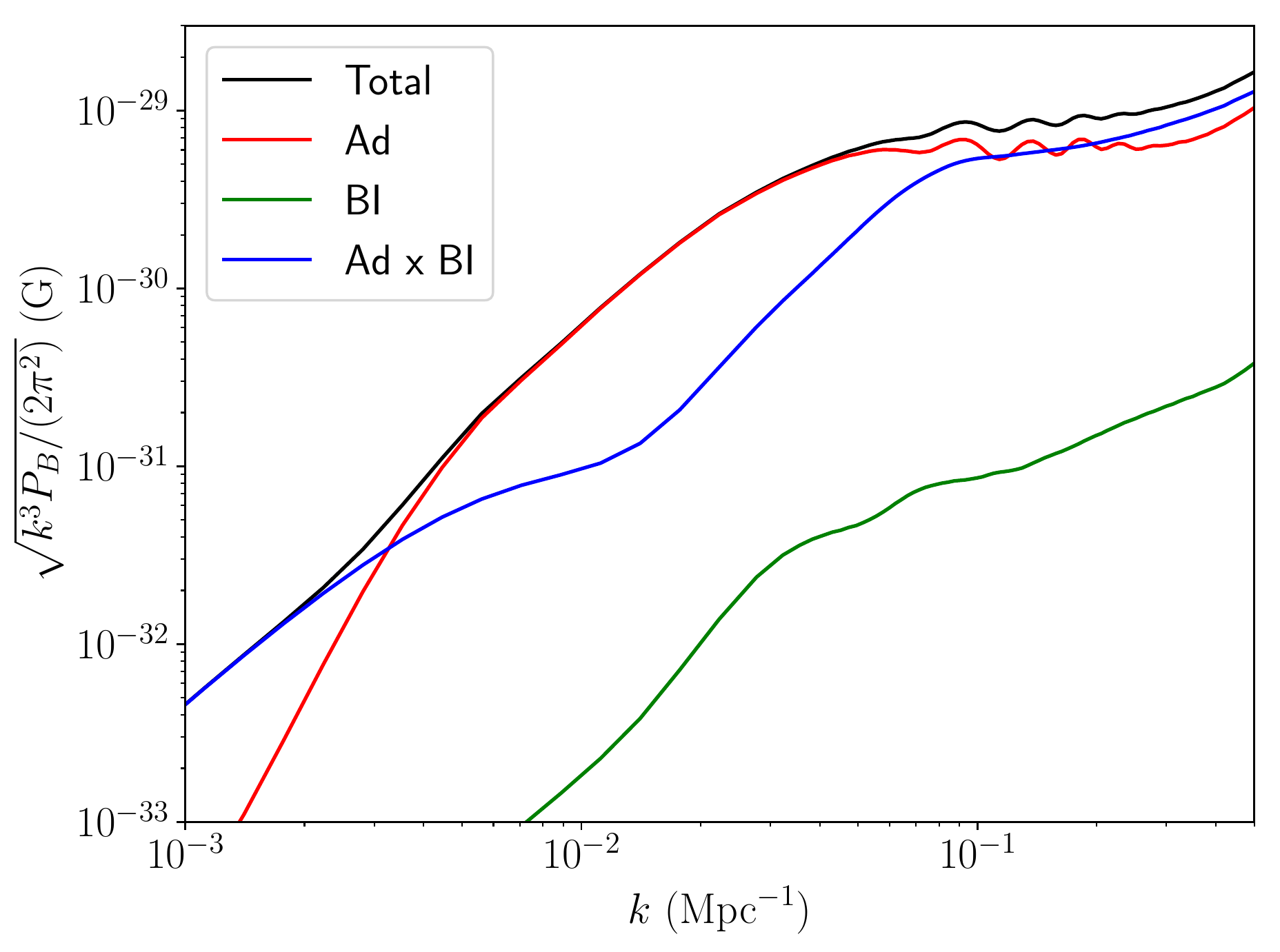}
    \includegraphics[width=0.49\textwidth]{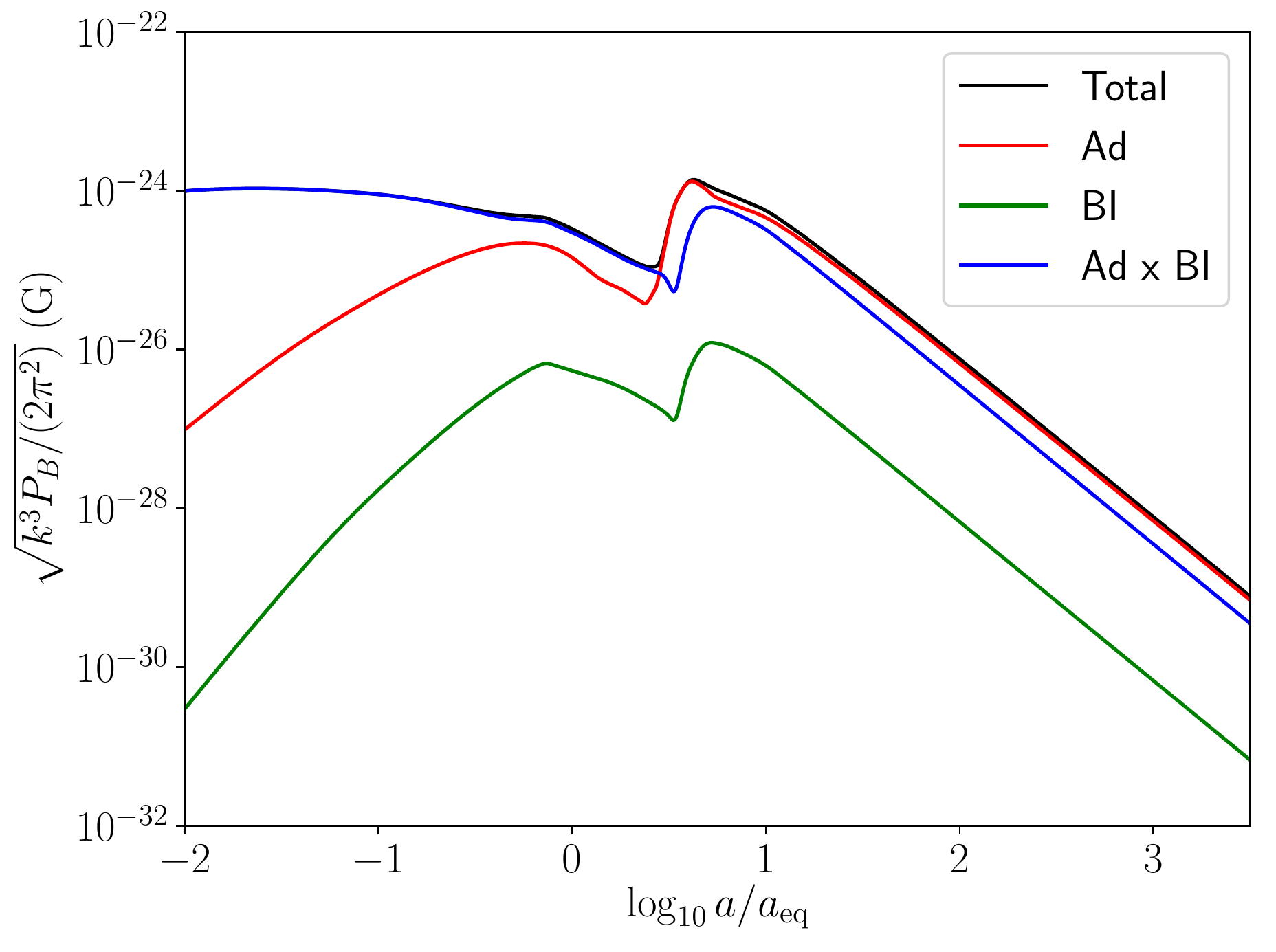}
    \caption{Normalized magnetic field power spectrum sourced by the mixture of adiabatic and baryon isocurvature modes as a function of wave number, $k$, at $z=0$ (left) and as a function of scale factor, $a$, for $k=0.06\ \text{Mpc}^{-1}$ (right).}
    \label{fig:ADBI}
\end{figure}

Figure~\ref{fig:ADBI} shows the results for the baryon isocurvature mode. We can see that the contribution from the pure baryon isocurvature mode (BI) is much smaller than the adiabatic mode (Ad). However, the contribution sourced by the mixture of the adiabatic and baryon isocurvature modes (Ad x BI) is similar in size to the adiabatic mode on Mpc scales and enhances the total result by approximately  60\%. On intermediate scales, the largest contribution is still the adiabatic one, but, on very large scales, the largest contribution is from the mixed mode once more. In particular, the spectral index of the mixed contribution on large scales differs by at least $\Delta n=1$ relative to that of the adiabatic mode. However, as we will see below, this is dependent on the spectral index of the isocurvature spectrum.

Regarding the time evolution shown on the right-hand plot of Fig.~\ref{fig:ADBI}, we see that both pure modes have a similar evolution, in spite of the amplitude difference. The mixed contribution shows substantial differences. In particular, at very early times, the mixed mode is several orders of magnitude larger than the pure modes. This is due to the effects described at the end of Section~\ref{sec:mag} and in Ref.~\cite{Maeda:2011uq}, which also explain the large contribution from the mixed mode on large scales. However, at late times, the mixed mode has a smaller contribution, even dropping below that of the adiabatic mode for this scale ($k=0.06\ \text{Mpc}^{-1}$). This clearly demonstrates that the results of Maeda et al.~\cite{Maeda:2011uq}, in which the evolution was not followed beyond matter-radiation equality, where too optimistic. It is clear from the case presented here, that, after recombination, the magnetic field generated by the adiabatic mode is comparable to that generated by the mixed mode, in spite of the latter being larger at all times prior to recombination.

\begin{figure}
    \centering
    \includegraphics[width=0.49\textwidth]{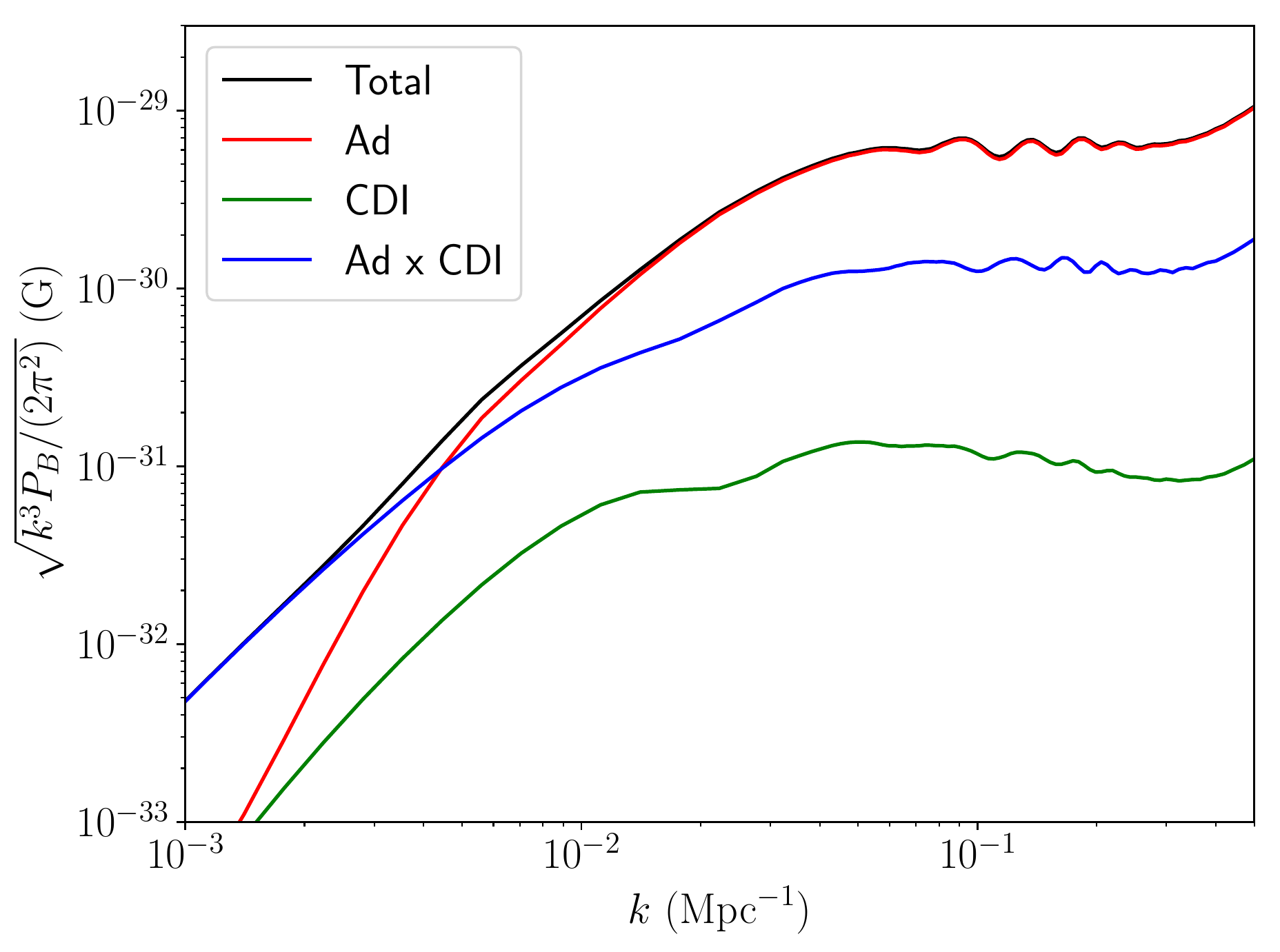}
    \includegraphics[width=0.49\textwidth]{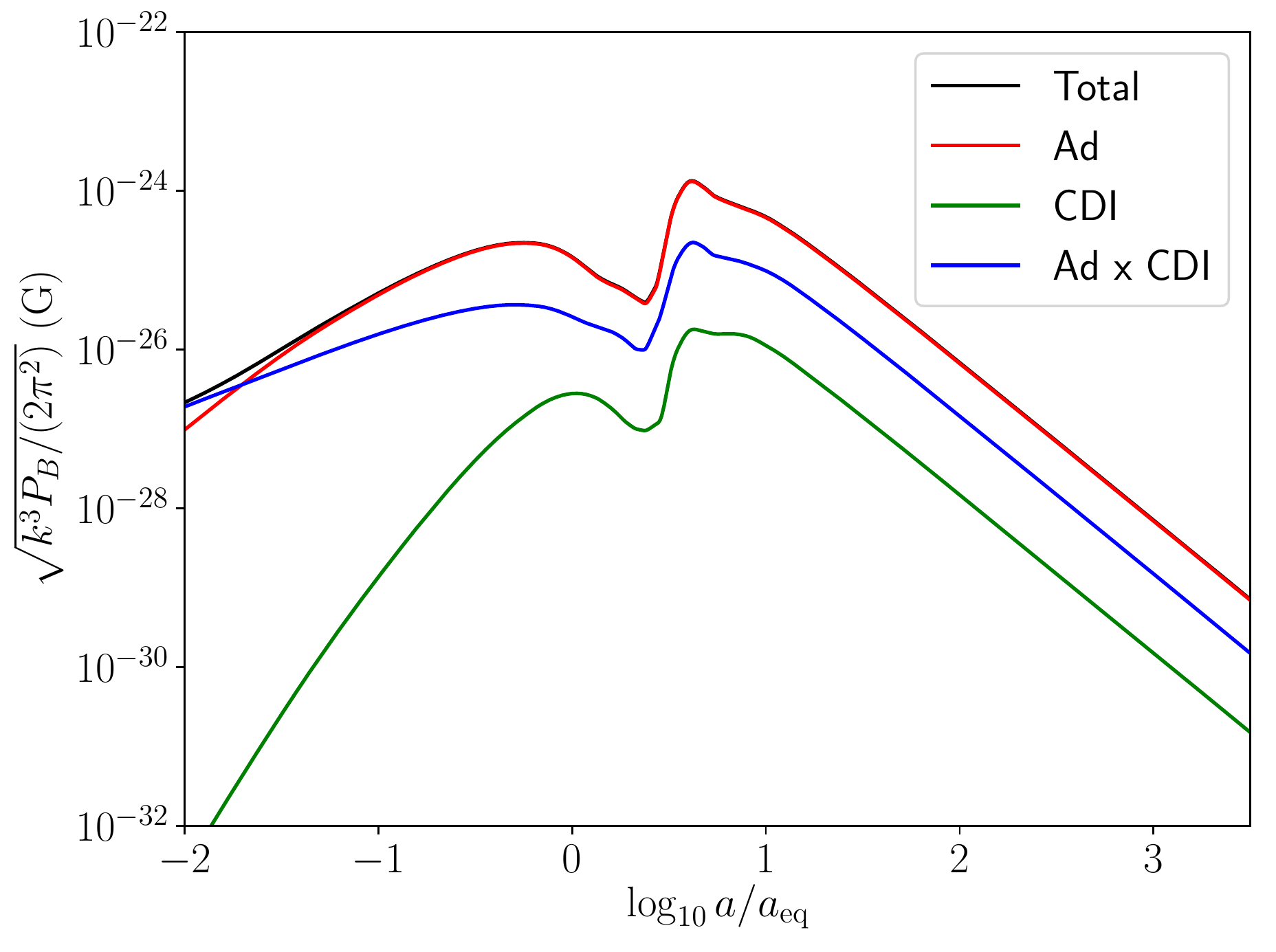}
    \caption{Normalized magnetic field power spectrum sourced by the mixture of adiabatic and cold dark matter isocurvature modes as a function of wave number, $k$, at $z=0$ (left) and as a function of scale factor, $a$, for $k=0.06\ \text{Mpc}^{-1}$ (right).}
    \label{fig:ADCDI}
\end{figure}

The result for the cold dark matter isocurvature mode is shown in Fig.~\ref{fig:ADCDI}. We can see that the contribution from the mixed mode (Ad x CDI) is much smaller in this case than it was when the baryon mode was active. This is not surprising, as dark matter does not play a role in magnetogenesis. In spite of the adiabatic mode dominating on small scales, the mixed mode still dominates on the largest scales, for the same reasons as described above --- all mixed modes generate a redder spectrum for the magnetic field on large scales than the pure modes. The time evolution of this solution also shows the mixed mode to be dominant at very early times, but with a smaller amplitude in this case.

\begin{figure}
    \centering
    \includegraphics[width=0.49\textwidth]{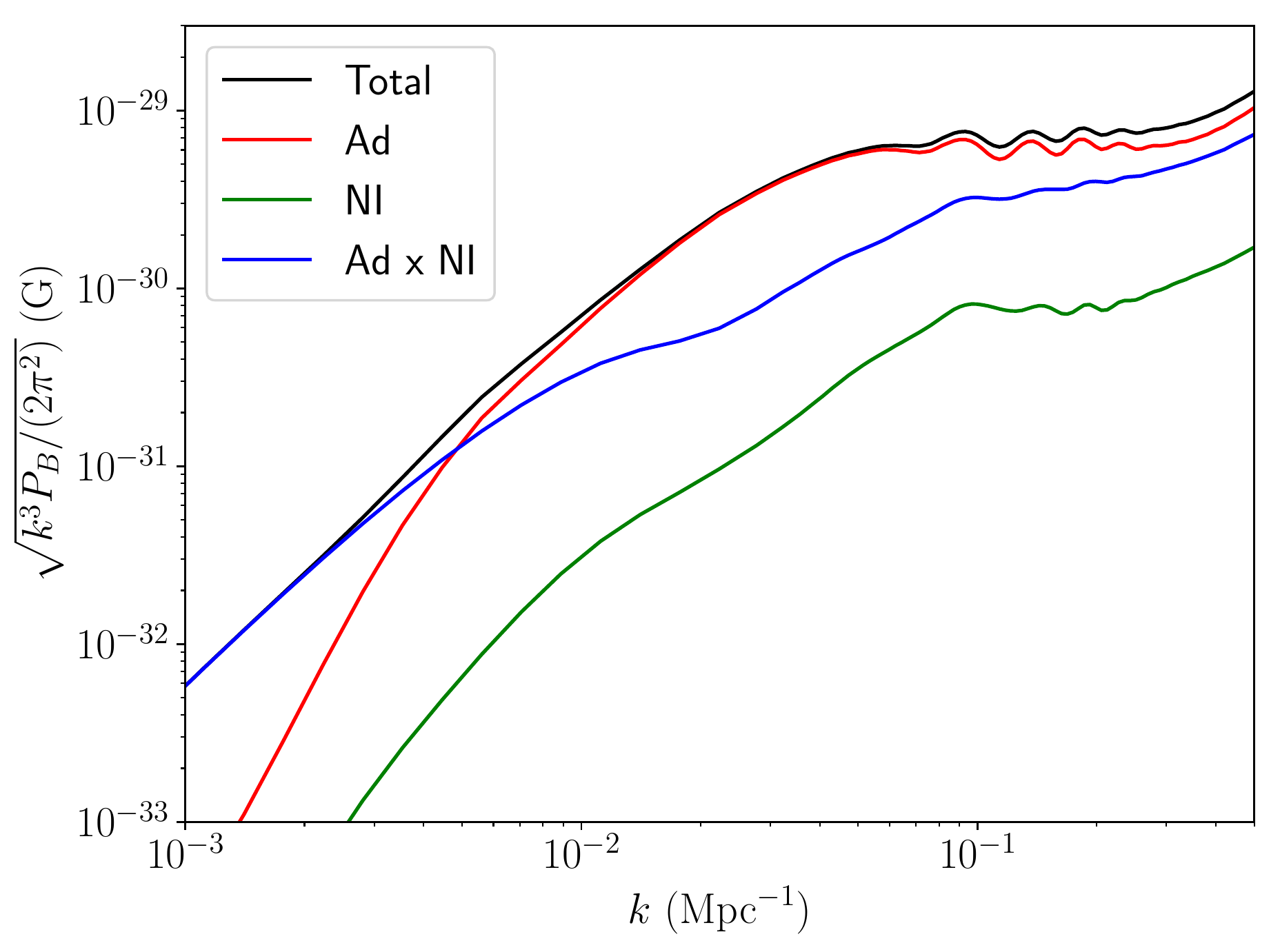}
    \includegraphics[width=0.49\textwidth]{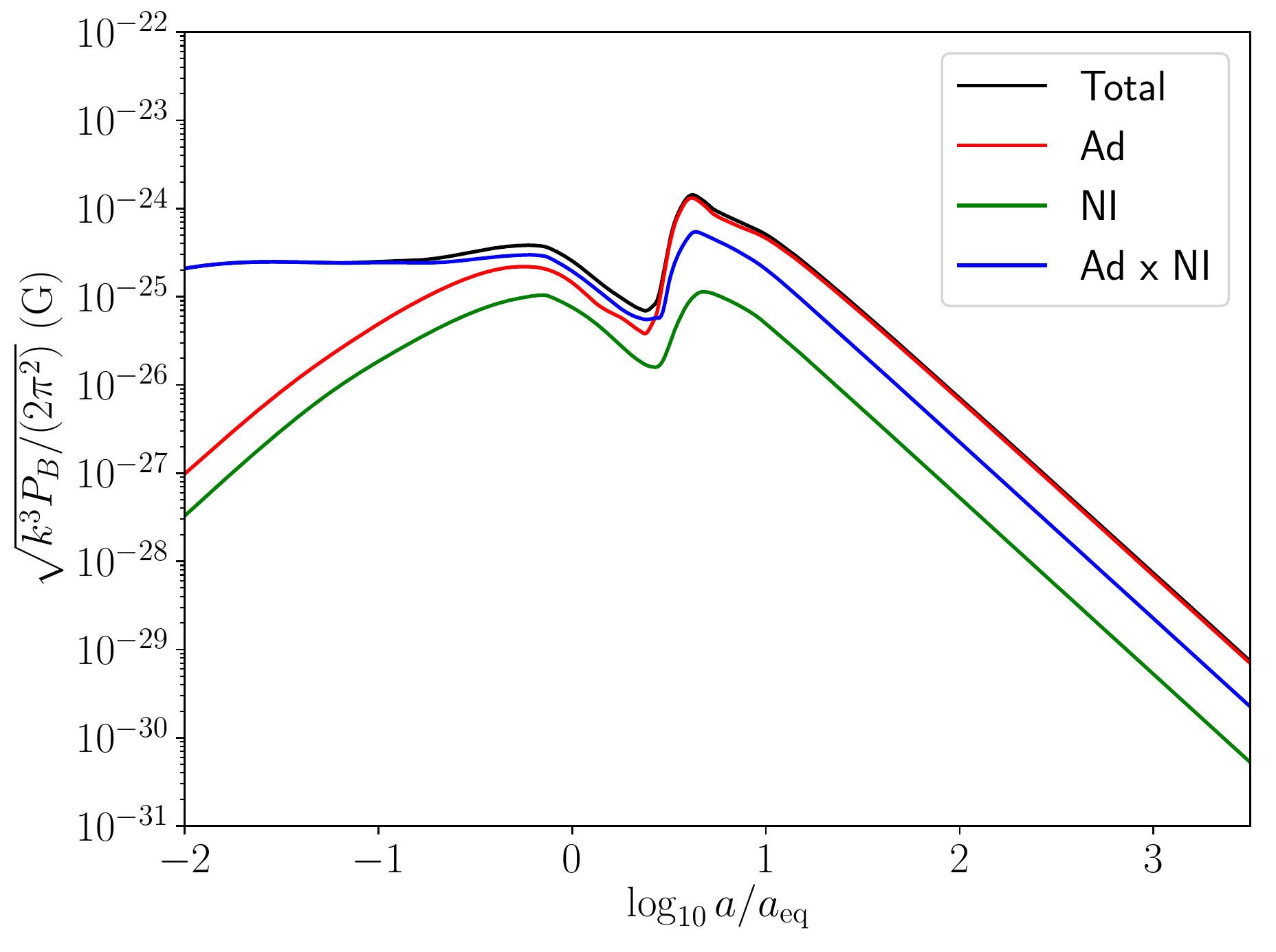}
    \caption{Normalized magnetic field power spectrum sourced by the mixture of adiabatic and neutrino isocurvature modes as a function of wave number, $k$, at $z=0$ (left) and as a function of scale factor, $a$, for $k=0.06\ \text{Mpc}^{-1}$ (right).}
    \label{fig:ADNI}
\end{figure}

Finally, we show the plots for the neutrino isocurvature mode in Fig.~\ref{fig:ADNI}. Similarly to the dark matter mode, the contributions from the neutrino mode are smaller on small scales than those generated by the adiabatic mode. However, the difference is now $O(1)$, instead of an order of magnitude as before, and for that reason, the total result is slightly enhanced on Mpc scales. On very large scales, and early times, the results are similar to the other modes.

We should note once more that all the results above depend crucially on the properties of the initial spectra. The dependence on the amplitudes $A_{aa}$ is simple to derive from Eq.~\eqref{SpcTra}. We note that, if the mixed mode dominates the isocurvature contribution, we have $P_B\propto A_{aa}$, while we have $P_B\propto A_{aa}^2$ if the pure mode is dominant. In all cases shown, the dominant contribution from isocurvatures is the mixed mode, and given that the amplitude of isocurvatures is not expected to be greater than $A_s$, we expect the dependence to be $P_B\propto A_{aa}$, in any realistic case. The upper bounds for the amplitudes of the isocurvature modes given by Planck are: $A_{\text{bb}}\approx 1.12 A_s$ for the baryon mode, $A_{\text{cc}}\approx 0.04 A_s$ for the dark matter mode and $A_{\nu\nu}\approx 0.07 A_s$ for the neutrino density isocurvature~\cite{Akrami:2018odb}. This implies that, when re-scaled to the appropriate amplitude, the contributions of both the dark matter and the neutrino mode to the magnetic field are almost negligible. However, the baryon isocurvature mode is allowed to increase slightly with respect to the plot, giving a result of $\sqrt{k^3 P_B/(2\pi^2)}=1.73\times10^{-29}\ \text{G}$ at $k=0.5\ \text{Mpc}^{-1}$, which should be compared to the result for the adiabatic mode of $1.05\times10^{-29}\ \text{G}$. This amounts to a relative enhancement of approximately $64\ \%$.

\begin{figure}
    \centering
    \includegraphics[width=0.49\textwidth]{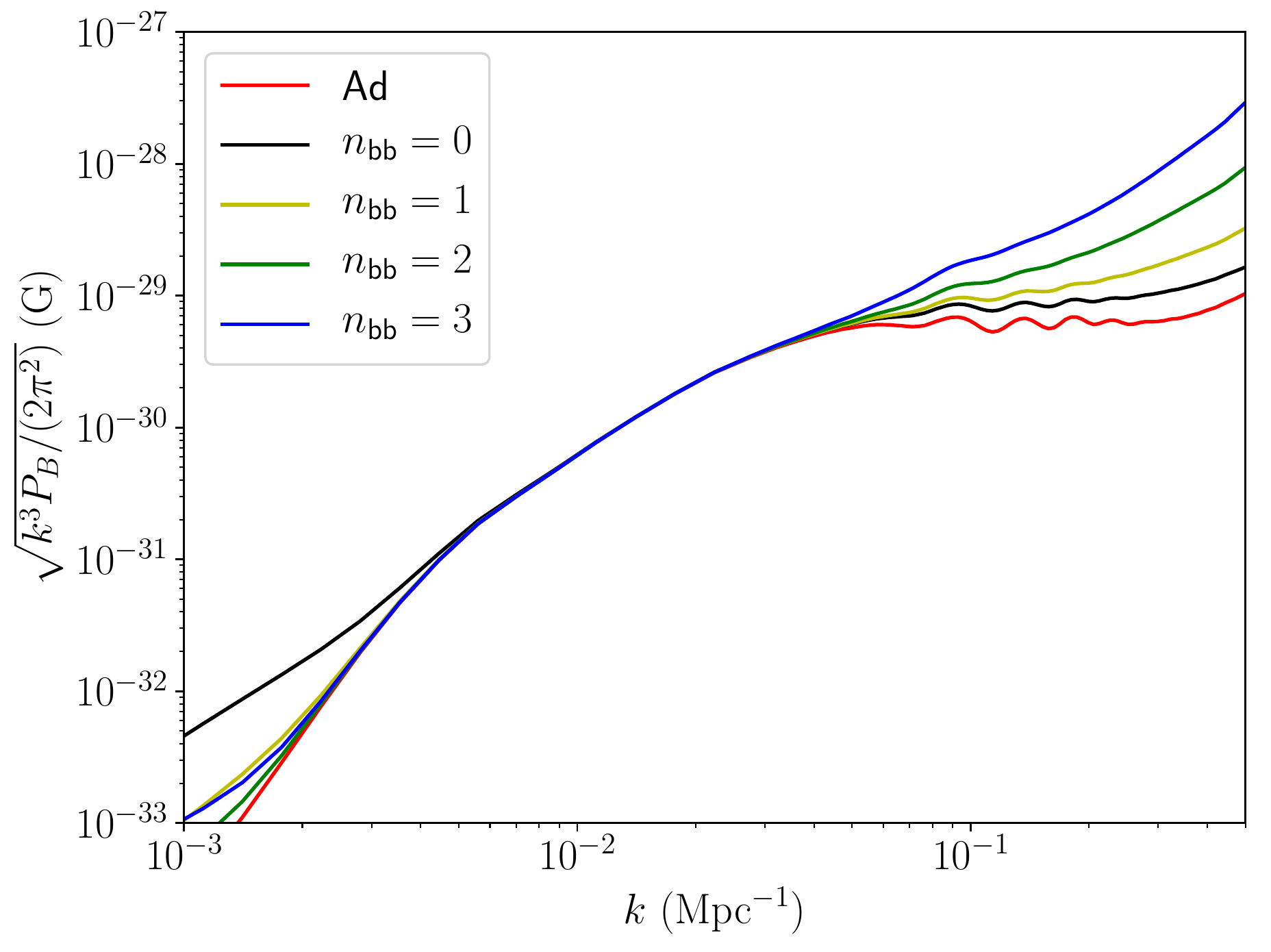}
    \includegraphics[width=0.49\textwidth]{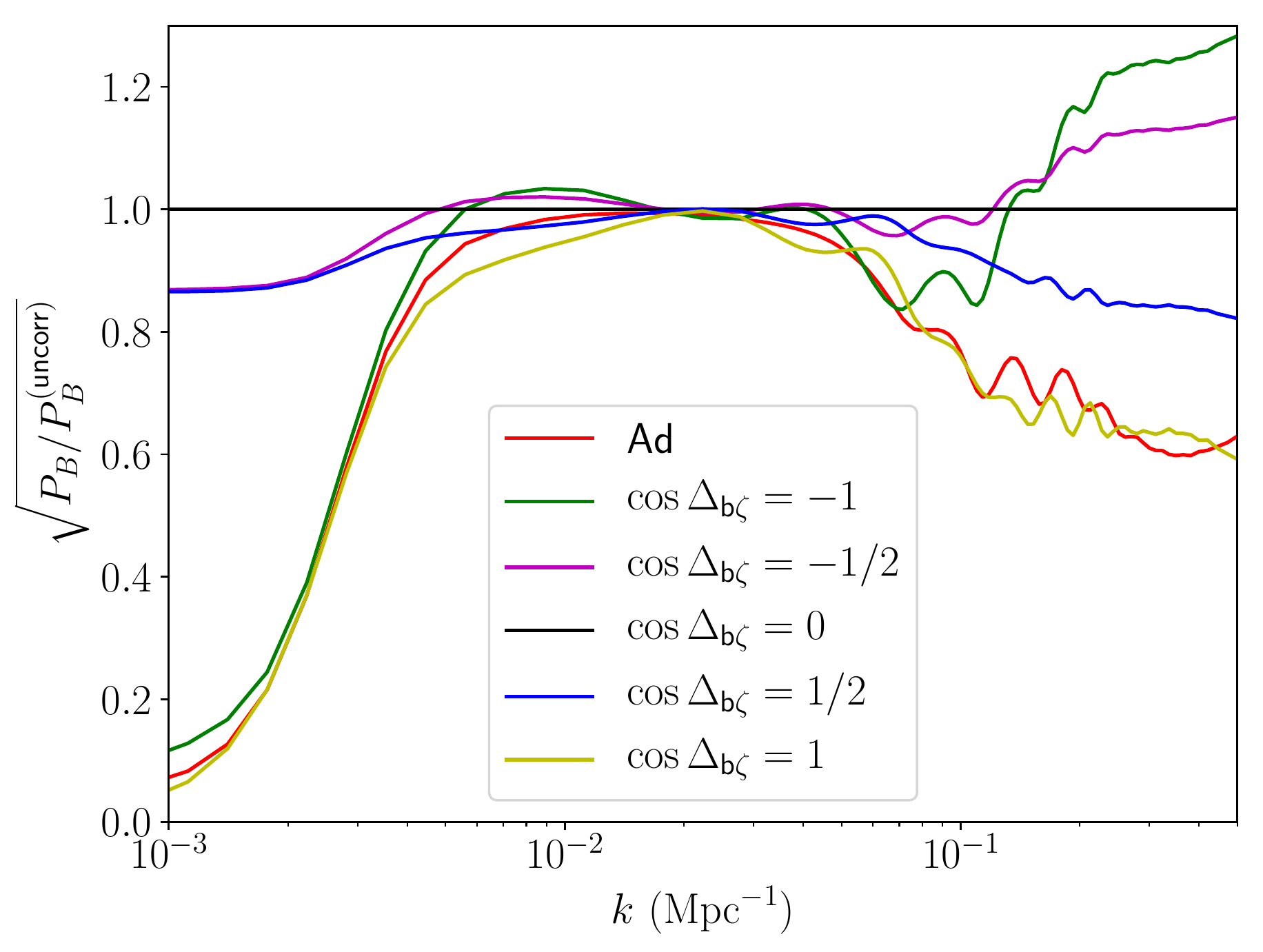}
    \caption{Normalized magnetic field power spectrum sourced by the mixture of adiabatic and baryon isocurvature modes as a function of wave number, $k$, at $z=0$ for different spectral indices of the baryon mode, $n_{\text{bb}}$ (left), and for different correlation angles, $\cos\Delta_{\text{b}\zeta}$ (right).}
    \label{fig:ADBIncos}
\end{figure}

The dependence on the spectral index can be seen on the left-hand side of Fig.~\ref{fig:ADBIncos}, for the baryon mode. We plot only the sum of all contributions, instead of the different mixtures of sourced modes. We see that on Mpc scales, the magnetic field is amplified further for bluer input spectra, with the opposite happening on very large scales. The case $n_{\text{bb}}=3$ is the one used by Maeda et al.~\cite{Maeda:2011uq} and results in a magnetic field which is approximately 30 times larger than the adiabatic case at $k=0.5\ \text{Mpc}^{-1}$ and with a bluer spectral index. The most recent Planck results prefer a spectral index for the baryon isocurvature mode that is closer to $n_{\text{bb}}=1$~\cite{Akrami:2018odb}, for which the enhancement factor of the magnetic field is only approximately 3.

The right-hand side of Fig.~\ref{fig:ADBIncos} shows how the resulting magnetic field varies when the adiabatic mode has a non-zero correlation with the baryon isocurvature mode. We note that partially and fully anti-correlated baryon isocurvatures give rise to a larger enhancement of the magnetic field at small scales, bringing the magnetic field produced to roughly double that generated with the adiabatic mode alone. A positive correlation always reduces the enhancement due to the presence of the isocurvature, even almost eliminating it for the fully correlated case. Additionally, we note that for both the fully correlated and anti-correlated cases, the behaviour approaches that of the adiabatic mode on large scales. This was already expected due to the arguments presented at the end of Section \ref{sec:mag} --- the curls of gradients of scalars are suppressed on large scales, if the scalars in question are correlated. One should note, however, that the  level of correlation currently allowed by Planck is small, with the reported 95\% CL interval being $\cos\Delta_{\text{b}\zeta}\in[-0.12,0.15]$ and these levels would not affect the magnetic field substantially.

The combination of both a blue spectral index ($n_{\text{bb}}=3$) and a full anti-correlation between the isocurvature and adiabatic modes was also tested and was found not to yield a substantially larger improvement, resulting in an enhancement ratio of approximately 31 instead of the 30 found with only $n_{\text{bb}}=3$. For other modes, similar enhancements can be found when varying the spectral index and the correlation angle. However, the baryon isocurvature mode is the one that gives rise to the largest magnetic field in all comparable cases.

We can conclude here that the presence of single isocurvature modes does enhance the magnetic field produced during the pre-recombination stage. However, this improvement is at most of an order of magnitude in the most extreme case, bringing the magnetic field to $3.3\times 10^{-28}$ G in the most optimistic scenario. This is still far from the levels required to explain observations ($\sim10^{-15}$ G).

\subsection{Magnetogenesis with compensated isocurvature mode}

We now move on to the case of the compensated isocurvature mode. This is a mode in which the density fluctuations in the non-relativistic matter species compensate each other so that no total matter entropy fluctuation exists. This is achieved by the relation
\be
\delta_{\text{b}}=-\frac{\Omega_{\text{c}}}{\Omega_{\text{b}}}\delta_{\text{c}}\,,
\ee
which is equivalent to 
\be
A_{\text{bb}}=\lb\frac{\Omega_{\text{c}}}{\Omega_{\text{b}}}\rb^2A_{\text{cc}}\,,\ \cos\Delta_{\text{bc}}=-1\,.
\ee
This mode is interesting because it has essentially no effect at the linear level. It has particularly little effect on the CMB, but can be constrained, as it can contribute to the smoothing of the peaks of the angular power spectrum of the CMB at small angular scales. This effect is similar to that generated by lensing and, for that reason, can help reduce the tension between different observables of the lensing potential. The most recent Planck results give a best-fit value for the amplitude of this mode of $2.2\substack{+1.0 \\ -1.5}\times 10^{-3}$, for the scale-invariant, uncorrelated case~\cite{Akrami:2018odb}, which is roughly $10^6$ times larger than the adiabatic mode, confirming similar results from other authors~\cite{Munoz:2015fdv,Valiviita:2017fbx}.

We have previously shown in Ref.~\cite{Carrilho:2018mqy} that this compensated mode can generate evolution of some second-order cosmological fluctuations at early times, when this mode is mixed with an adiabatic mode. It is thus expected that this mode will generate effects at second order, which may be measurable, given the large possible amplitude of this mode. 

\begin{figure}
    \centering
    \includegraphics[width=0.49\textwidth]{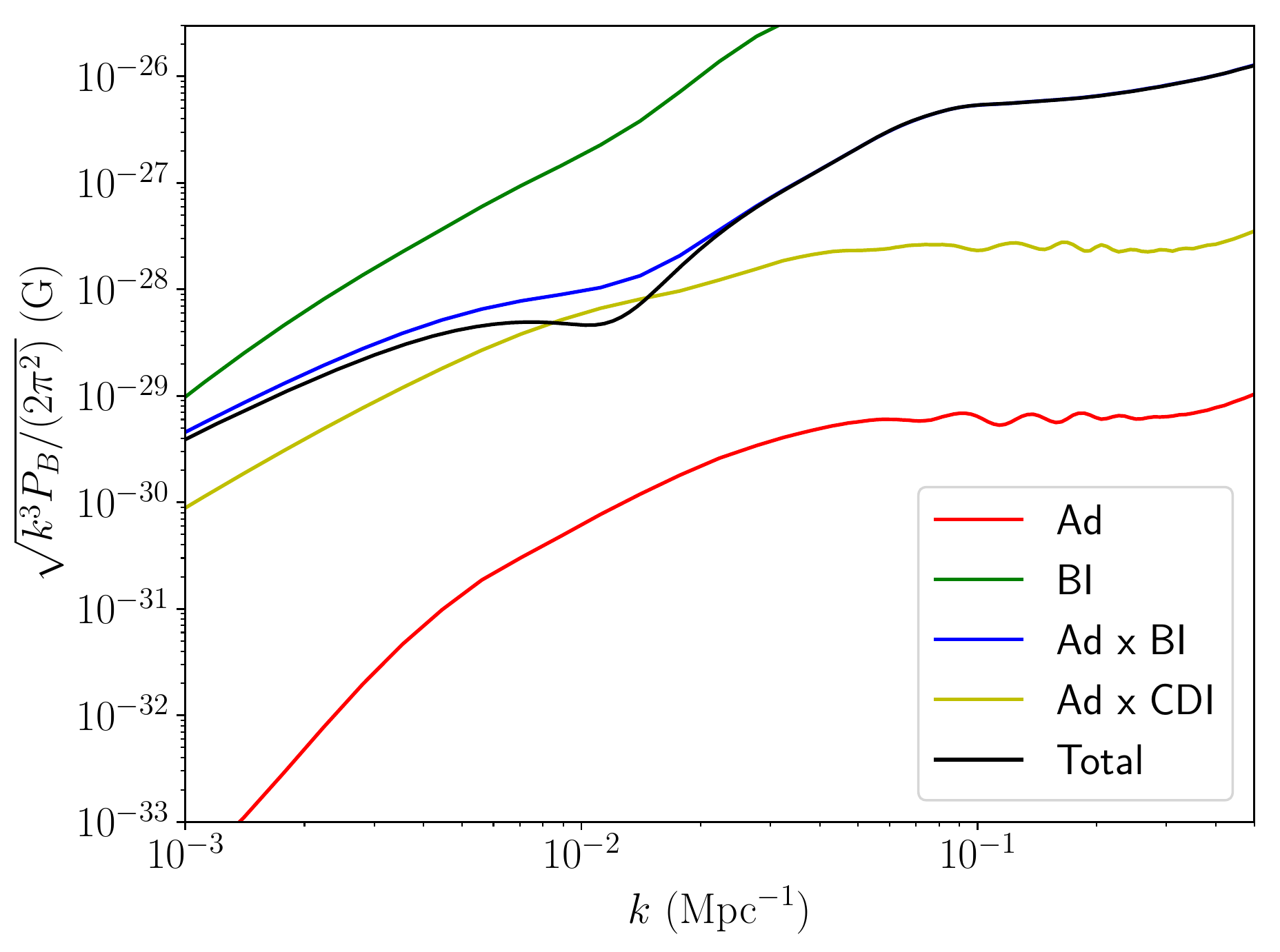}
    \includegraphics[width=0.49\textwidth]{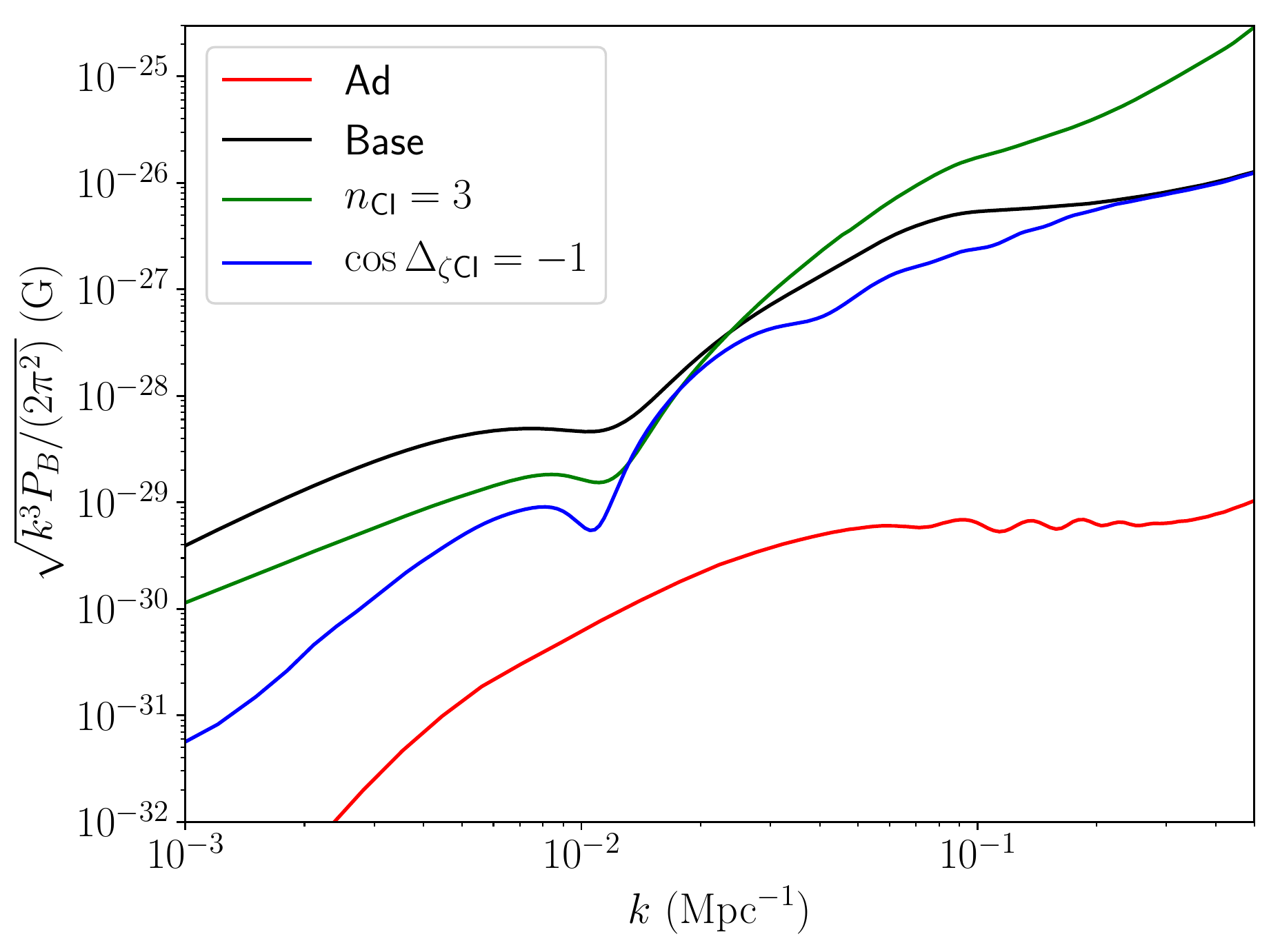}
    \caption{Normalized magnetic field power spectrum sourced by the mixture of adiabatic and compensated isocurvature modes as a function of wave number, $k$, at $z=0$, showing the contributions from different sources for the uncorrelated, scale-invariant case (left) and showing the effects of a non-zero spectral index of the compensated mode, $n_{\text{CI}}$ and of a non-zero correlation angle, $\cos\Delta_{\zeta\text{CI}}$ (right).}
    \label{fig:ADCIP}
\end{figure}

Figure~\ref{fig:ADCIP} shows our results for the magnetic field generated by this compensated mode, with the amplitude quoted above. In our notation we approximate this to
\be
A_{\text{bb}}=10^6 A_s\,,\ A_{\text{cc}}=(186.417)^2 A_s\,.
\ee
The most striking result is that the magnetic field is amplified by more than $10^3$ with respect to the purely adiabatic case on Mpc scales. Analysing Figs.~\ref{fig:ADBI} and \ref{fig:ADCDI}, we note that both the shapes and sizes of the contributions from the baryon and dark matter modes are different, so it is not so surprising to find that they do not cancel out when mixed together, even in this anti-correlated case. We see also that this result is dominated by the contribution from the mixed baryon-adiabatic mode, in spite of the pure baryon isocurvature mode being considerably larger. This is were the compensation between the two matter modes has an effect, with the two pure modes canceling each other out. Furthermore, this implies that the magnetic field power spectrum is proportional to the amplitude of the baryon isocurvature mode, thus explaining why the plotted magnetic field ($\propto \sqrt{P_B}$) is approximately $10^3$ times larger than that shown in Fig.~\ref{fig:ADBI}.

The plot on the right-hand-side of Fig.~\ref{fig:ADCIP} shows the effects of changing the spectral index of the compensated mode and its correlation angle with the adiabatic mode. We note that, contrarily to what occurred for the baryon isocurvature contribution, the anti-correlation between the adiabatic and compensated modes does not give rise to an increased magnetic field at any scale in the range studied here. However, we see that the effect of a blue spectral index is similar to that shown before, enhancing the magnetic field by a factor of approximately 30 on small scales, bringing the magnetic field at $k=0.5\ \text{Mpc}^{-1}$  to $3\times10^{-25}$ G. It should me noted, though, that the compensated mode used by Planck~\cite{Akrami:2018odb} was the scale-invariant, uncorrelated mode, a fact that was important for the lensing degeneracy, and it is not clear how alternative scenarios would affect that. However, a fully correlated mode is better motivated from the theoretical point of view, as it can be generated in curvaton models of the early Universe and would, in principle, be easier to detect~\cite{He:2015msa}. Models with non-scale-invariant spectra have not been explored. They may be harder to constrain for blue spectra, as the current methods rely on a large-scale modulation of the baryon-dark matter fraction, which would require larger amplitudes to be detected with a blue spectral index.

We see from these results that a compensated isocurvature mode can generate considerably larger magnetic fields than the adiabatic mode, given its potential large amplitude. In our most optimistic scenario, we see that the magnetic field can be enhanced up to 4 orders of magnitude on Mpc scales, which certainly improves the prospects of this magnetic field being detected in the future.

\section{Conclusions}
\label{sec:conc}

We have computed the magnetic field generated in the pre-recombination epoch in the presence of isocurvature modes. We have shown how the different initial conditions can affect the magnetic field power spectrum and concluded that it is the non-linear mode mixing that provides the greatest contributions. In particular, the baryon isocurvature mode, when mixed with the adiabatic mode, has the greatest potential to enhance the magnetic field. We have also demonstrated that isocurvature modes with bluer spectral indices give rise to larger improvements than those of the scale-invariant case, resulting in a magnetic field approximately 30 times larger than with the adiabatic mode, for the most optimistic case. Furthermore, we have also seen that a variation of the correlation fraction between the adiabatic mode and the isocurvature mode in question can modify the results, slightly enhancing the Mpc-scale magnetic field in the fully anti-correlated case and suppressing it in the opposite limit.

The compensated isocurvature mode can have a much larger effect on the magnetic field, since its amplitude is not as constrained as the others. We find that the uncorrelated, scale-invariant compensated mode can source a magnetic field of order $10^{-26}$ G, 3 orders of magnitude larger than the adiabatic case. For very blue input spectra this can once again be improved by a factor of 30, reaching $3\times10^{-25}$ G at $k=0.5\ \text{Mpc}^{-1}$. While possibly not sufficient to explain the tentative observations of void magnetic fields of $10^{-15}$ G, this is certainly a step in the right direction. Furthermore, the mechanisms for generating isocurvature fluctuations are often the source of primordial non-Gaussianity, whose contribution to the magnetic field could also be appreciable, given its dependence on the four-point function of primordial fluctuations. Further enhancements may be possible in the presence of features in the primordial spectra, particularly in the compensated isocurvature case. Both of these conjectured contributions could play an important role in magnetic field generation and should be further investigated.

This result regarding the compensated isocurvature mode could have further consequences. Note that a mode with such a large amplitude has virtually no effect on the CMB, but does leave a huge imprint on the magnetic field. This indicates that other quantities that appear only at second order in cosmological perturbations could be similarly enhanced by this mode, which would allow us to better constrain a primordial compensated isocurvature fluctuation. We shall study some of those quantities in future publications, including gravitational waves generated at second order in fluctuations, as well as the bispectrum of the CMB.

\acknowledgments

PC is supported by STFC grant ST/P000592/1 and by the Funda\c{c}\~{a}o para a Ci\^{e}ncia e Tecnologia (FCT) grant SFRH/BD/118740/2016. KAM is supported, in part, by STFC grant ST/P000592/1.
The tensor algebra package xAct \cite{xAct}\footnote{\href{http://www.xact.es}{http://www.xact.es}}, as well as its
sub-package xPand \cite{xPand}\footnote{\href{http://www.xact.es/xPand/}{http://www.xact.es/xPand/}}, were used in the derivation of
many of the equations presented in this work.
The authors are grateful to David Mulryne for useful discussions.

\appendix
\section{Initial conditions}
\label{sec:ini}

In this appendix, we present the initial conditions used in SONG for the divergence-free vector variables. These are approximate solutions for the evolution equations, written as a power series in the conformal time, $\tau$. They are found using the same techniques as in Ref.~\cite{Carrilho:2018mqy}. We show here only the results for Poisson gauge, as that is the one used in SONG. We show the results for the curls of the vector potential, $S_i$, the vector parts of the velocities of all species and the vector part of the anisotropic stress of neutrinos, which we describe using the variable $\tilde\sigma_\nu^i\equiv-\Pi_\nu^i/2\rho_\nu $. We show the curls of variables, as those give rise to real transfer functions in Fourier space. For the magnetic field, we show instead $a B_{\text{v}}^i$, as it is proportional to a curl. In all cases, we show the Fourier transform of the variable in question and shorten our notation by showing only the second-order transfer function multiplied by the initial conditions, thus omitting the integrals over the momenta (as are present, for example in Eq.~\eqref{Btransfer}). The variables identifying the initial conditions are shown here as $\psi_{k_i}^0$, $\delta_{\text{b},k_i}^0$, $\delta_{\text{c},k_i}^0$ and $\delta_{\nu,k_i}^0$, which should be evaluated in synchronous gauge, as explained in Ref.~\cite{Carrilho:2018mqy}. They are equivalent to the set $\lbc\zeta,S_{\text{c}\gamma},S_{\text{b}\gamma},S_{\nu\gamma}\rbc$. In the results shown below, we use a set of ratios to simplify our notation. They are 
\begin{align}
\omega\equiv\frac{\Omega_M \mathcal{H}}{\sqrt{\Omega_R}}\,,\ 
R_{\text{c}}\equiv\frac{\Omega_{\text{c}}}{\Omega_M}\,,\ R_{\text{b}}\equiv\frac{\Omega_{\text{b}}}{\Omega_M}\,,\ R_\nu\equiv\frac{\Omega_\nu}{\Omega_R}\,,\ R_\gamma\equiv\frac{\Omega_\gamma}{\Omega_R}\,,
\end{align}
where the $\Omega_s$ are the usual density parameters of all species, as well as those for the total matter ($M$) and radiation ($R$).

\subsection{Pure adiabatic mode}

\begin{align}
    &\n\times \itbf{S}^{(2)}=\n\times \itbf{v}_{\text{c}}^{(2)}=\n\times \itbf{v}_{\text{b}\gamma}^{(2)}=\n\times \itbf{v}_\nu^{(2)}=-\frac{40(5+R_\nu)\lb k_1^2-k_2^2\rb}{(15+4 R_\nu)^2k^2}\itbf{k}_1\times \itbf{k}_2\tau\,\psi^{0}_{k_1}\psi^{0}_{k_2}\,,\nonumber\\
    &\n\times \tilde{\boldsymbol{\sigma}}_\nu^{(2)}=\frac{4\lb k_1^2-k_2^2\rb}{3(15+4 R_\nu)k^2}\itbf{k}_1\times \itbf{k}_2\tau^2\,\psi^{0}_{k_1}\psi^{0}_{k_2}\,,\\
    &a\itbf{B}^{(2)}_{\text{v}}=\frac{m_p}{e}\frac{\lb k_1^2-k_2^2\rb\itbf{k}_1\times \itbf{k}_2\tau^3}{252(25+2 R_\nu)(15+4 R_\nu)^4}f_{\psi}(R_\nu)\,\psi^{0}_{k_1}\psi^{0}_{k_2}\,,\nonumber
\end{align}
with
\be
f_\psi(R_\nu)=2953125+696250 R_\nu-1013000R_\nu^2-503200R_\nu^3-76160 R_\nu^4-3584 R_\nu^5\,.\nonumber
\ee

\subsection{Pure baryon isocurvature mode}

\begin{align}
    &\n\times \itbf{S}^{(2)}=15R_{\text{b}}^2\omega^2\frac{(-225+110R_\nu+16R_\nu^2)\lb k_1^2-k_2^2\rb}{16(15+2 R_\nu)^2(25+2 R_\nu)k^2}\itbf{k}_1\times \itbf{k}_2\tau^3\,\delta^{0}_{\text{b},k_1}\delta^{0}_{\text{b},k_2}\,,\nonumber\\
    &\n\times \itbf{v}_{\text{c}}^{(2)}=\n\times \itbf{v}_{\text{b}\gamma}^{(2)}=\n\times \itbf{v}_\nu^{(2)}=15R_{\text{b}}^2\omega^2\frac{(-225+110R_\nu+16R_\nu^2)\lb k_1^2-k_2^2\rb}{16(15+2 R_\nu)^2(25+2 R_\nu)k^2}\itbf{k}_1\times \itbf{k}_2\tau^3\,\delta^{0}_{\text{b},k_1}\delta^{0}_{\text{b},k_2}\,,\nonumber\\
    &\n\times \tilde{\boldsymbol{\sigma}}_\nu^{(2)}=R_{\text{b}}^2\omega^2\frac{(-825-70R_\nu+4R_\nu^2)\lb k_1^2-k_2^2\rb}{12(15+2 R_\nu)^2(25+2 R_\nu)k^2}\itbf{k}_1\times \itbf{k}_2\tau^4\,\delta^{0}_{\text{b},k_1}\delta^{0}_{\text{b},k_2}\,,\\
    &a\itbf{B}^{(2)}_{\text{v}}=O(\tau^4)\,.\nonumber
\end{align}

\subsection{Mixture of adiabatic and baryon isocurvature modes}

\begin{align}
    &\n\times \itbf{S}^{(2)}=15R_{\text{b}}\omega\frac{5\lb k_1^2-k_2^2\rb+R_\nu\lb k^2-k_1^2+k_2^2\rb}{2(15+2 R_\nu)(15+4 R_\nu)k^2}\itbf{k}_1\times \itbf{k}_2\tau^2\,\delta^{0}_{\text{b},k_1}\psi^{0}_{k_2}\,,\nonumber\\
    &\n\times \itbf{v}_{\text{c}}^{(2)}=-R_{\text{b}}\omega\frac{90(R_\nu-5)\lb k_1^2-k_2^2\rb+\lb 75-50 R_\nu+16 R_\nu^2\rb k^2}{12(15+2 R_\nu)(15+4 R_\nu)k^2}\itbf{k}_1\times \itbf{k}_2\tau^2\,\delta^{0}_{\text{b},k_1}\psi^{0}_{k_2}\,,\nonumber\\
    &\n\times \itbf{v}_{\text{b}\gamma}^{(2)}=\n\times \itbf{v}_\nu^{(2)}=-5R_{\text{b}}\omega\frac{3(R_\nu-5)\lb k_1^2-k_2^2\rb-\lb 15+8 R_\nu\rb k^2}{2(15+2 R_\nu)(15+4 R_\nu)k^2}\itbf{k}_1\times \itbf{k}_2\tau^2\,\delta^{0}_{\text{b},k_1}\psi^{0}_{k_2}\,,\nonumber\\
    &\n\times \tilde{\boldsymbol{\sigma}}_\nu^{(2)}=-R_{\text{b}}\omega\frac{(4R_\nu-5)\lb k_1^2-k_2^2\rb+15 k^2}{3(15+2 R_\nu)(15+4 R_\nu)k^2}\itbf{k}_1\times \itbf{k}_2\tau^3\,\delta^{0}_{\text{b},k_1}\psi^{0}_{k_2}\,,\\
    &a\itbf{B}^{(2)}_{\text{v}}=\frac{m_p}{e}R_{\text{b}}\omega\frac{375-700 R_\nu-320 R_\nu^2-32 R_\nu^3}{12(15+2 R_\nu)(15+4 R_\nu)^2}\itbf{k}_1\times \itbf{k}_2\tau^2\,\delta^{0}_{\text{b},k_1}\psi^{0}_{k_2}\,.\nonumber
\end{align}

\subsection{Pure cold dark matter isocurvature mode}

\begin{align}
    &\n\times \itbf{S}^{(2)}=15R_{\text{c}}^2\omega^2\frac{(-225+110R_\nu+16R_\nu^2)\lb k_1^2-k_2^2\rb}{16(15+2 R_\nu)^2(25+2 R_\nu)k^2}\itbf{k}_1\times \itbf{k}_2\tau^3\,\delta^{0}_{\text{c},k_1}\delta^{0}_{\text{c},k_2}\,,\nonumber\\
    &\n\times \itbf{v}_{\text{c}}^{(2)}=\n\times \itbf{v}_{\text{b}\gamma}^{(2)}=\n\times \itbf{v}_\nu^{(2)}=15R_{\text{c}}^2\omega^2\frac{(-225+110R_\nu+16R_\nu^2)\lb k_1^2-k_2^2\rb}{16(15+2 R_\nu)^2(25+2 R_\nu)k^2}\itbf{k}_1\times \itbf{k}_2\tau^3\,\delta^{0}_{\text{c},k_1}\delta^{0}_{\text{c},k_2}\,,\nonumber\\
    &\n\times \tilde{\boldsymbol{\sigma}}_\nu^{(2)}=R_{\text{c}}^2\omega^2\frac{(-825-70R_\nu+4R_\nu^2)\lb k_1^2-k_2^2\rb}{12(15+2 R_\nu)^2(25+2 R_\nu)k^2}\itbf{k}_1\times \itbf{k}_2\tau^4\,\delta^{0}_{\text{c},k_1}\delta^{0}_{\text{c},k_2}\,,\\
    &a\itbf{B}^{(2)}_{\text{v}}=O(\tau^4)\,.\nonumber
\end{align}

\subsection{Mixture of adiabatic and cold dark matter isocurvature modes}

\begin{align}
    &\n\times \itbf{S}^{(2)}=15R_{\text{c}}\omega\frac{5\lb k_1^2-k_2^2\rb+R_\nu\lb k^2-k_1^2+k_2^2\rb}{2(15+2 R_\nu)(15+4 R_\nu)k^2}\itbf{k}_1\times \itbf{k}_2\tau^2\,\delta^{0}_{\text{c},k_1}\psi^{0}_{k_2}\,,\nonumber\\
    &\n\times \itbf{v}_{\text{c}}^{(2)}=-R_{\text{c}}\omega\frac{90(R_\nu-5)\lb k_1^2-k_2^2\rb+\lb 75-50 R_\nu+16 R_\nu^2\rb k^2}{12(15+2 R_\nu)(15+4 R_\nu)k^2}\itbf{k}_1\times \itbf{k}_2\tau^2\,\delta^{0}_{\text{c},k_1}\psi^{0}_{k_2}\,,\nonumber\\
    &\n\times \itbf{v}_{\text{b}\gamma}^{(2)}=\n\times \itbf{v}_\nu^{(2)}=-5R_{\text{c}}\omega\frac{3(R_\nu-5)\lb k_1^2-k_2^2\rb-\lb 15+8 R_\nu\rb k^2}{2(15+2 R_\nu)(15+4 R_\nu)k^2}\itbf{k}_1\times \itbf{k}_2\tau^2\,\delta^{0}_{\text{c},k_1}\psi^{0}_{k_2}\,,\nonumber\\
    &\n\times \tilde{\boldsymbol{\sigma}}_\nu^{(2)}=-R_{\text{c}}\omega\frac{(4R_\nu-5)\lb k_1^2-k_2^2\rb+15 k^2}{3(15+2 R_\nu)(15+4 R_\nu)k^2}\itbf{k}_1\times \itbf{k}_2\tau^3\,\delta^{0}_{\text{c},k_1}\psi^{0}_{k_2}\,,\\
    &a\itbf{B}^{(2)}_{\text{v}}=\frac{m_p}{e}R_{\text{c}}\omega\frac{375-700 R_\nu-320 R_\nu^2-32 R_\nu^3}{12(15+2 R_\nu)(15+4 R_\nu)^2}\itbf{k}_1\times \itbf{k}_2\tau^2\,\delta^{0}_{\text{c},k_1}\psi^{0}_{k_2}\,.\nonumber
\end{align}

\subsection{Mixture of baryon and cold dark matter isocurvature modes}

\begin{align}
    &\n\times \itbf{S}^{(2)}=15R_{\text{c}}R_{\text{b}}\omega^2\frac{(-225+110R_\nu+16R_\nu^2)\lb k_1^2-k_2^2\rb}{16(15+2 R_\nu)^2(25+2 R_\nu)k^2}\itbf{k}_1\times \itbf{k}_2\tau^3\,\delta^{0}_{\text{b},k_1}\delta^{0}_{\text{c},k_2}\,,\nonumber\\
    &\n\times \itbf{v}_{\text{c}}^{(2)}=\n\times \itbf{v}_\nu^{(2)}=15R_{\text{c}}R_{\text{b}}\omega^2\frac{(-225+110R_\nu+16R_\nu^2)\lb k_1^2-k_2^2\rb}{16(15+2 R_\nu)^2(25+2 R_\nu)k^2}\itbf{k}_1\times \itbf{k}_2\tau^3\,\delta^{0}_{\text{b},k_1}\delta^{0}_{\text{c},k_2}\,,\nonumber\\
    &\n\times \itbf{v}_{\text{b}\gamma}^{(2)}=15R_{\text{c}} R_{\text{b}}\omega^2f_{cb}(k,k_1,k_2, R_\nu)\itbf{k}_1\times \itbf{k}_2\tau^3\,\delta^{0}_{\text{b},k_1}\delta^{0}_{\text{c},k_2}\,,\nonumber\\
    &\n\times \tilde{\boldsymbol{\sigma}}_\nu^{(2)}=R_{\text{c}}R_{\text{b}}\omega^2\frac{(-825-70R_\nu+4R_\nu^2)\lb k_1^2-k_2^2\rb}{12(15+2 R_\nu)^2(25+2 R_\nu)k^2}\itbf{k}_1\times \itbf{k}_2\tau^4\,\delta^{0}_{\text{b},k_1}\delta^{0}_{\text{c},k_2}\,,\\
    &a\itbf{B}^{(2)}_{\text{v}}=\frac{m_p}{e}\frac{R_{\text{c}}R_{\text{b}}}{16R_\gamma}\omega^2\itbf{k}_1\times \itbf{k}_2\tau^3\,\delta^{0}_{\text{b},k_1}\delta^{0}_{\text{c},k_2}\,,\nonumber
\end{align}

with
\be
f_{cb}(k,k_1,k_2, R_\nu)=\frac{(15+2 R_\nu)^2(25+2 R_\nu)k^2-15(225-335R_\nu+94R_\nu^2+16R_\nu^3)\lb k_1^2-k_2^2\rb}{16R_\gamma(15+2 R_\nu)^2(25+2 R_\nu)k^2}\,.\nonumber
\ee

\subsection{Pure neutrino isocurvature mode}

\begin{align}
    &\n\times \itbf{S}^{(2)}=\n\times \itbf{v}_{\text{c}}^{(2)}=\n\times \itbf{v}_{\text{b}\gamma}^{(2)}=\n\times \itbf{v}_\nu^{(2)}=-\frac{95R_\nu^2\lb k_1^2-k_2^2\rb}{2R_\gamma(15+4 R_\nu)^2k^2}\itbf{k}_1\times \itbf{k}_2\tau\,\delta^{0}_{\nu,k_1}\delta^{0}_{\nu,k_2}\,,\nonumber\\
    &\n\times \tilde{\boldsymbol{\sigma}}_\nu^{(2)}=-\frac{\lb 225-153 R_\nu+4R_\nu^2\rb\lb k_1^2-k_2^2\rb}{12R_\gamma(15+4 R_\nu)k^2}\itbf{k}_1\times \itbf{k}_2\tau^2\,\delta^{0}_{\nu,k_1}\delta^{0}_{\nu,k_2}\,,\\
    &a\itbf{B}^{(2)}_{\text{v}}=-\frac{m_p}{e}\frac{R_\nu^2\lb k_1^2-k_2^2\rb\itbf{k}_1\times \itbf{k}_2\tau^3}{1008R_\gamma^2(25+2 R_\nu)(15+4 R_\nu)^4}f_{\nu}(R_\nu)\,\delta^{0}_{\nu,k_1}\delta^{0}_{\nu,k_2}\,,\nonumber
\end{align}

with
\be
f_{\nu}(R_\nu)= 12487500 +    R_\nu (12220875 +   8 R_\nu (360420 +  R_\nu (-10387 + 8 R_\nu (-221 + 28 R_\nu))))\,.\nonumber
\ee

\subsection{Mixture of adiabatic and neutrino isocurvature modes}

\begin{align}
    &\n\times \itbf{S}^{(2)}=-10R_\nu\frac{ 2 k^2-3k_1^2+3k_2^2}{(15+4 R_\nu)^2k^2}\itbf{k}_1\times \itbf{k}_2\tau\,\delta^{0}_{\nu,k_1}\psi^{0}_{k_2}\,,\nonumber\\
    &\n\times \itbf{v}_{\text{c}}^{(2)}=2R_\nu\frac{15\lb k_1^2-k_2^2\rb+\lb 5+4 R_\nu\rb k^2}{(15+4 R_\nu)^2k^2}\itbf{k}_1\times \itbf{k}_2\tau\,\delta^{0}_{\nu,k_1}\psi^{0}_{k_2}\,,\nonumber\\
    &\n\times \itbf{v}_{\text{b}\gamma}^{(2)}=R_\nu\frac{60R_\gamma\lb k_1^2-k_2^2\rb+\lb 245+116 R_\nu\rb k^2}{2R_\gamma(15+4 R_\nu)^2k^2}\itbf{k}_1\times \itbf{k}_2\tau\,\delta^{0}_{\nu,k_1}\psi^{0}_{k_2}\,,\nonumber\\
    &\n\times \itbf{v}_\nu^{(2)}=5\frac{12R_\nu\lb k_1^2-k_2^2\rb-5\lb 9+4 R_\nu\rb k^2}{(15+4 R_\nu)^2k^2}\itbf{k}_1\times \itbf{k}_2\tau\,\delta^{0}_{\nu,k_1}\psi^{0}_{k_2}\,,\nonumber\\
    &\n\times \tilde{\boldsymbol{\sigma}}_\nu^{(2)}=\frac{\lb 5+4 R_\nu\rb\lb k_1^2-k_2^2\rb+10 k^2}{(15+4 R_\nu)^2k^2}\itbf{k}_1\times \itbf{k}_2\tau\,\delta^{0}_{\nu,k_1}\psi^{0}_{k_2}\,,\\
    &a\itbf{B}^{(2)}_{\text{v}}=-\frac{m_p}{e}R_\nu\frac{1425+520 R_\nu+176 R_\nu^2+64 R_\nu^3}{4R_\gamma(15+4 R_\nu)^3}\itbf{k}_1\times \itbf{k}_2\tau\,\delta^{0}_{\nu,k_1}\psi^{0}_{k_2}\,.\nonumber
\end{align}

\subsection{Mixture of baryon and neutrino isocurvature modes}

\begin{align}
    &\n\times \itbf{S}^{(2)}=15R_{\text{b}}R_\nu\omega\frac{(2R_\nu-15)k^2-(135+22R_\nu)\lb k_1^2-k_2^2\rb}{8(15+2 R_\nu)^2(15+4 R_\nu)k^2}\itbf{k}_1\times \itbf{k}_2\tau^2\,\delta^{0}_{\nu,k_1}\delta^{0}_{\text{b},k_2}\,,\nonumber\\
    &\n\times \itbf{v}_{\text{c}}^{(2)}=-R_{\text{b}}R_\nu\omega\frac{(32R_\nu^2+390R_\nu+2475)k^2+45(135+22R_\nu)\lb k_1^2-k_2^2\rb}{24(15+2 R_\nu)^2(15+4 R_\nu)k^2}\itbf{k}_1\times \itbf{k}_2\tau^2\,\delta^{0}_{\nu,k_1}\delta^{0}_{\text{b},k_2}\,,\nonumber\\
    &\n\times \itbf{v}_\nu^{(2)}=15R_{\text{b}}\omega\frac{(8R_\nu^2+30R_\nu+225)k^2-2R_\nu(135+22R_\nu)\lb k_1^2-k_2^2\rb}{16(15+2 R_\nu)^2(15+4 R_\nu)k^2}\itbf{k}_1\times \itbf{k}_2\tau^2\,\delta^{0}_{\nu,k_1}\delta^{0}_{\text{b},k_2}\,,\nonumber\\
    &\n\times \itbf{v}_{\text{b}\gamma}^{(2)}=R_{\text{b}}R_\nu\omega f_{b\nu}(k,k_1,k_2, R_\nu)\itbf{k}_1\times \itbf{k}_2\tau^2\,\delta^{0}_{\nu,k_1}\delta^{0}_{\text{b},k_2}\,,\\
    &\n\times \tilde{\boldsymbol{\sigma}}_\nu^{(2)}=-R_{\text{b}}\omega\frac{15(2R_\nu-15)k^2-(1125+150R_\nu-8R_\nu^2)\lb k_1^2-k_2^2\rb}{12(15+2 R_\nu)^2(15+4 R_\nu)k^2}\itbf{k}_1\times \itbf{k}_2\tau^3\,\delta^{0}_{\nu,k_1}\delta^{0}_{\text{b},k_2}\,,\nonumber\\
    &a\itbf{B}^{(2)}_{\text{v}}=-R_\nu R_{\text{b}}\frac{m_p}{e}\frac{23400+29175 R_\nu+6174 R_\nu^2+328 R_\nu^3+32 R_\nu^4}{48R_\gamma^2(15+4R_\nu)^2(15+2R_\nu)}\omega\itbf{k}_1\times \itbf{k}_2\tau^2\,\delta^{0}_{\nu,k_1}\delta^{0}_{\text{b},k_2}\,,\nonumber
\end{align}

with
\be
f_{b\nu}(k,k_1,k_2, R_\nu)=\frac{(88R_\nu^3-406R_\nu^2-1305R_\nu-14850)k^2-30R_\gamma^2(135+22R_\nu)\lb k_1^2-k_2^2\rb}{16R_\gamma^2(15+2 R_\nu)^2(15+4 R_\nu)k^2}\,.\nonumber
\ee

\subsection{Mixture of cold dark matter and neutrino isocurvature modes}

\begin{align}
    &\n\times \itbf{S}^{(2)}=15R_{\text{c}} R_\nu\omega\frac{(2R_\nu-15)k^2-(135+22R_\nu)\lb k_1^2-k_2^2\rb}{8(15+2 R_\nu)^2(15+4 R_\nu)k^2}\itbf{k}_1\times \itbf{k}_2\tau^2\,\delta^{0}_{\nu,k_1}\delta^{0}_{\text{c},k_2}\,,\nonumber\\
    &\n\times \itbf{v}_{\text{c}}^{(2)}=-R_{\text{c}} R_\nu\omega\frac{(32R_\nu^2+390R_\nu+2475)k^2+45(135+22R_\nu)\lb k_1^2-k_2^2\rb}{24(15+2 R_\nu)^2(15+4 R_\nu)k^2}\itbf{k}_1\times \itbf{k}_2\tau^2\,\delta^{0}_{\nu,k_1}\delta^{0}_{\text{c},k_2}\,,\nonumber\\
    &\n\times \itbf{v}_\nu^{(2)}=15R_{\text{c}}\omega\frac{(8R_\nu^2+30R_\nu+225)k^2-2R_\nu(135+22R_\nu)\lb k_1^2-k_2^2\rb}{16(15+2 R_\nu)^2(15+4 R_\nu)k^2}\itbf{k}_1\times \itbf{k}_2\tau^2\,\delta^{0}_{\nu,k_1}\delta^{0}_{\text{c},k_2}\,,\nonumber\\
    &\n\times \itbf{v}_{\text{b}\gamma}^{(2)}=R_{\text{c}} R_\nu\omega f_{c\nu}(k,k_1,k_2, R_\nu)\itbf{k}_1\times \itbf{k}_2\tau^2\,\delta^{0}_{\nu,k_1}\delta^{0}_{\text{c},k_2}\,,\\
    &\n\times \tilde{\boldsymbol{\sigma}}_\nu^{(2)}=-R_\text{c}\omega\frac{15(2R_\nu-15)k^2-(1125+150R_\nu-8R_\nu^2)\lb k_1^2-k_2^2\rb}{12(15+2 R_\nu)^2(15+4 R_\nu)k^2}\itbf{k}_1\times \itbf{k}_2\tau^3\,\delta^{0}_{\nu,k_1}\delta^{0}_{\text{c},k_2}\,,\nonumber\\
    &a\itbf{B}^{(2)}_{\text{v}}=R_\nu R_{\text{c}}\frac{m_p}{e}\frac{6975+750 R_\nu-24 R_\nu^2+32 R_\nu^3}{48R_\gamma(15+4R_\nu)^2(15+2R_\nu)}\omega\itbf{k}_1\times \itbf{k}_2\tau^2\,\delta^{0}_{\nu,k_1}\delta^{0}_{\text{c},k_2}\,,\nonumber
\end{align}

with
\be
f_{c\nu}(k,k_1,k_2, R_\nu)=-\frac{(136R_\nu^2+630R_\nu+4725)k^2+30R_\gamma(135+22R_\nu)\lb k_1^2-k_2^2\rb}{16R_\gamma(15+2 R_\nu)^2(15+4 R_\nu)k^2}\,.\nonumber
\ee

\bibliographystyle{JHEPmodplain}
\bibliography{biblio_mag}

\end{document}